\pdfoutput=1
\documentclass[journal]{IEEEtran}
\usepackage{flushend}
\usepackage[latin1]{inputenc}
\usepackage{times,amsmath}
\usepackage{amsfonts}
\usepackage{subfigure}
\usepackage{multirow}
\usepackage{multicol}
\usepackage{enumerate}
\usepackage{graphicx}
\usepackage{mathtools, cuted}
\usepackage{MnSymbol}
\usepackage{stfloats}
\usepackage{subcaption}
\usepackage[table]{xcolor}
\usepackage[square, comma, sort&compress, numbers]{natbib}
\usepackage{nohyperref}
\usepackage{algorithm,algorithmic}
\usepackage{bigints}
\usepackage{amsmath}
\usepackage{nomencl}
\usepackage{tikz}
\usetikzlibrary{calc}

\tikzset{%
    every node/.style={draw=blue, minimum width=2cm, minimum height=1cm, fill=blue!50}
}
\newcounter{tempEquationCounter}
\newcounter{thisEquationNumber}

\makeatletter
\newcommand{\vast}{\bBigg@{4}}
\newcommand{\Vast}{\bBigg@{5}}
\newcommand\numeq[1]%
  {\stackrel{\scriptscriptstyle(\mkern-1.5mu#1\mkern-1.5mu)}{=}}
\makeatother

\graphicspath{ {Figures/} }
\begin{document}

\makenomenclature
\title{Integrating Communication, Sensing, and Security: Progress and Prospects of PLS in ISAC Systems }

\author{
Waqas Aman,~\IEEEmembership{Member,~IEEE,}
Elmehdi Illi,~\IEEEmembership{Member,~IEEE,} Saif Al-Kuwari,~\IEEEmembership{Senior Member,~IEEE}, and Marwa Qaraqe,~\IEEEmembership{Senior Member,~IEEE}
\thanks{ W. Aman, E. Illi, M. Qaraqe, and S. Al-Kuwari are with College of Science and Engineering, Hamad Bin Khalifa University, Doha, Qatar.
 Emails: \{waman, eilli, smalkuwari, mqaraqe\}@hbku.edu.qa.}}
\vspace{-0.8cm}

\maketitle

\begin{abstract}

The sixth generation of wireless networks defined several key performance indicators (KPIs) for assessing its networks, mainly in terms of reliability, coverage, and sensing. In this regard, remarkable attention has been paid recently to the integrated sensing and communication (ISAC) paradigm as an enabler for efficiently and jointly performing communication and sensing using the same spectrum and hardware resources. On the other hand, ensuring communication and data security has been an imperative requirement for wireless networks throughout their evolution. The physical-layer security (PLS) concept paved the way to catering to the security needs in wireless networks in a sustainable way while guaranteeing theoretically secure transmissions, independently of the computational capacity of adversaries. Therefore, it is of paramount importance to consider a balanced trade-off between communication reliability, sensing, and security in future networks, such as the 5G and beyond and the 6G. 
In this paper, we provide a comprehensive and system-wise review of designed secure ISAC systems from a PLS point of view. In particular, the impact of various physical-layer techniques, schemes, and wireless technologies to ensure the sensing-security trade-off is studied from the surveyed work. Furthermore, the amalgamation of PLS and ISAC is analyzed in a broader impact by considering attacks targeting data confidentiality, communication covertness, and sensing spoofing. The paper also serves as a tutorial by presenting several theoretical foundations on ISAC and PLS, which representing a practical guide for readers to develop novel secure ISAC network designs. 

\end{abstract}


\addcontentsline{toc}{section}{Nomenclature}

   \nomenclature{5G}{Fifth Generation}
   \nomenclature{5GB}{Fifth Generation and Beyond}
    \nomenclature{6G}{Sixth Generation}
    \nomenclature{AI}{Artificial Intelligence}
    \nomenclature{AN}{Artificial Noise}
    \nomenclature{AO}{Alternating Optimization}
    \nomenclature{AoA}{Angle of Arrival}
    \nomenclature{BCD}{Block Coordinate Descent}
    \nomenclature{BCRB}{Bayesian Cramer-Rao Bound}
    \nomenclature{BS}{Base Station}
    \nomenclature{CNNs}{Convolutional Neural Networks}
    \nomenclature{CSI}{Channel State Information}
    \nomenclature{CU}{Communication User}
    \nomenclature{DFRC}{Dual Function Radar Communication}
    \nomenclature{DL}{Downlink}
    \nomenclature{DRL}{Deep Reinforcement Learning}
    \nomenclature{EKF}{Extended Kalman Filter}
    \nomenclature{eMBB}{enhanced Mobile Broadband}
    \nomenclature{FAP}{False Alarm Probability}
    \nomenclature{MDP}{Missed Detection Probability}
    \nomenclature{FC-RIS}{Fully Connected 
    Reconfigurable Intelligent Surface}
    \nomenclature{FD}{Full Duplex}
     \nomenclature{FDMA}{Frequency Division Multiple Access}
    \nomenclature{FP}{Fractional Programming}
    \nomenclature{GEO}{Geostationary Earth Orbit}
    \nomenclature{HAPS}{High Altitude Platform Systems}
    \nomenclature{IoT}{Internet of Things}
    \nomenclature{IOS}{Intelligent Omni-Surfaces}
    \nomenclature{IRM}{Iterative Rank Minimization}
    \nomenclature{ISAC}{Integrated Sensing and Communication}
    \nomenclature{ISMR}{Integrated Sidelobe-to-Mainlobe Ratio}
     \nomenclature{KPIs}{Ket Performance Indicators}
    \nomenclature{LEO}{Low Earth Orbit}
    \nomenclature{LoS}{Line of Sight}
    \nomenclature{LPI}{Low Probability of Intercept}
    \nomenclature{LRT}{Likelihood Ratio Test}
    \nomenclature{MIMO}{Multiple-Input Multiple-Output}
    \nomenclature{MISO}{Multiple-Input Single-Output}
    \nomenclature{MM}{Majorization-Minimization}
    \nomenclature{ML}{Machine Learning}
    \nomenclature{MRT}{Maximal-Ratio Transmission}
      \nomenclature{MSE}{Mean-Squared Error}
    \nomenclature{NOMA}{Non-Orthogonal Multiple Access}
    \nomenclature{NTC}{Non-Terrestrial Communication}
     \nomenclature{OFDM}{Orthogonal Frequency Division Multiplexing}
     \nomenclature{OMA}{Orthogonal Multiple Access}
     \nomenclature{OTFS}{Orthogonal Time Frequency Space}
    \nomenclature{PCRB}{Posterior Cramer-Rao Bound}
    \nomenclature{PHY}{Physical Layer}
    \nomenclature{PLS}{Physical Layer Security}
    \nomenclature{Qos}{Quality of Service}
    \nomenclature{RCC}{Radar-Communication Coexistence}
    \nomenclature{RCS}{Radar Cross Section}
    \nomenclature{REM}{Radio Environment Mapping}
    \nomenclature{RFF}{Radio-Frequency Fingerprinting}
    \nomenclature{RIS}{Reconfigurable Intelligent Surface}
    \nomenclature{SCA}{Successive Convex Approximation}
    \nomenclature{SDR}{Semi-Definite Relaxation}
    \nomenclature{SIC}{Successive
Interference Cancellation}
    \nomenclature{SINR}{Signal-to-Interference-plus-Noise Ratio}
    \nomenclature{SNR}{Signal-to-Noise Ratio}
    \nomenclature{SOCP}{Second-Order Cone Programming}
    \nomenclature{SOP}{Secrecy Outage Probability}
    \nomenclature{STAR-RIS}{Simultaneously Transmitting and Reflecting Reconfigurable Intelligent Surface}
     \nomenclature{SRM}{Secrecy Rate Maximization}
    \nomenclature{TDMA}{Time Division Multiple Access}
    \nomenclature{THz}{Terahertz}
    \nomenclature{UAV}{Unmanned Aerial Vehicle}
    \nomenclature{UL}{Uplink}
    \nomenclature{ZSRP}{Zero Secrecy Rate Probability}

\printnomenclature
\section{Introduction}

The rapid evolution of wireless communication networks, from the fifth generation (5G) to the emerging sixth generation (6G), has catalyzed the development of innovative technologies that push the boundaries of connectivity, reliability, and sensing capabilities \cite{lin20243gpp}. In the 6G realm, several key performance indicators (KPIs) have been defined with target levels to achieve, such as a $99.999999\%$ of data reliability, a peak data rate of $1$ Tbps, and $10$ million device per squared meter in terms of connection density \cite{survey}. Furthermore, sensing is expected to play a pivotal role in various applications of 5G and beyond (5GB) and 6G, such as in augmented/virtual/extended reality (AR/VR/XR), remote surgery, autonomous vehicles, industrial automation, etc, where the wireless community's aim is set to reach a centimeter-level of sensing accuracy. 

\begin{figure*}
     \centering
     \includegraphics[width=\linewidth]{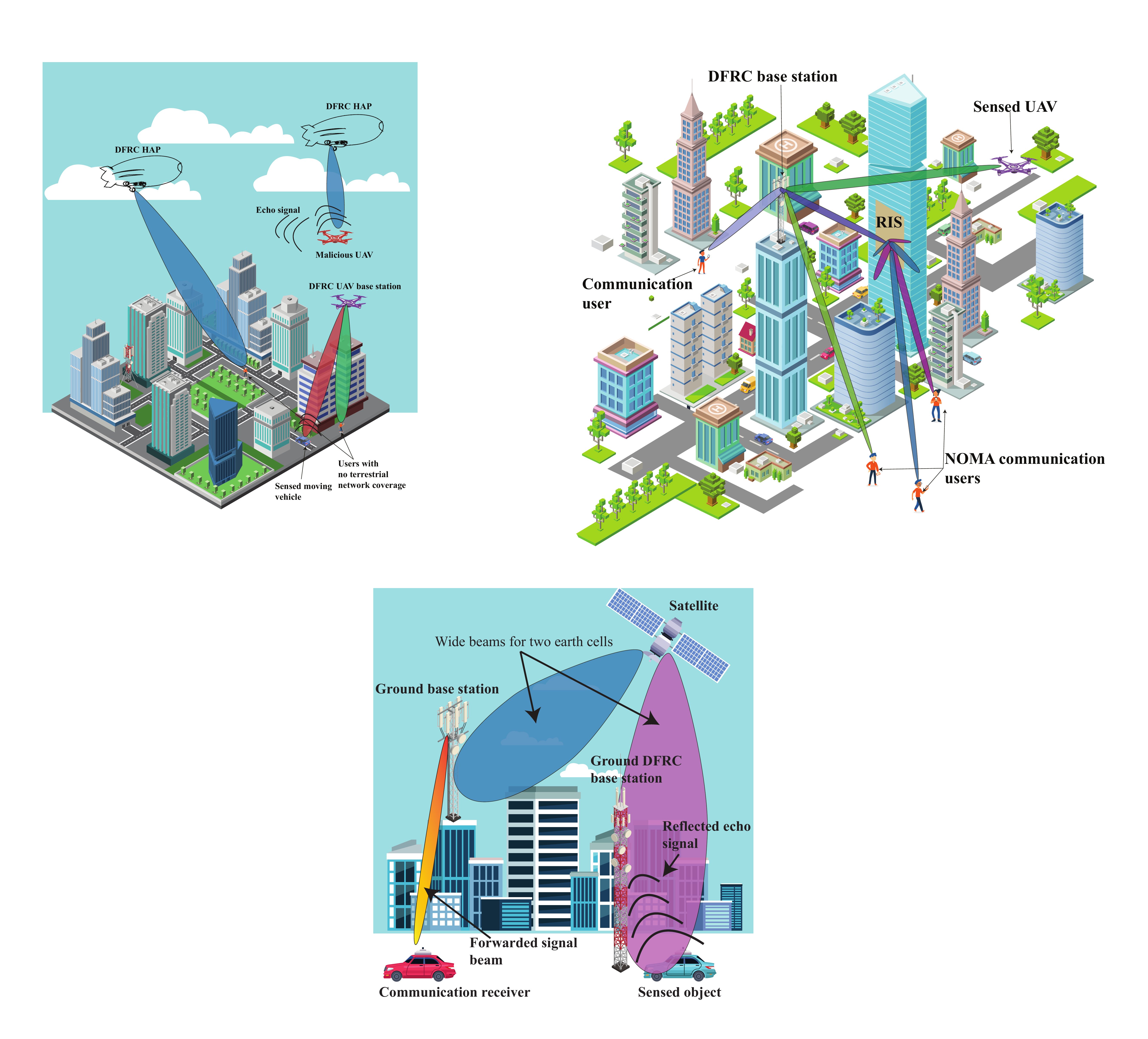}
     \caption{An illustration of the interplay of ISAC use-cases with cutting-edge wireless technologies and techniques.}
     \label{fig:ISAC_systems}
 \end{figure*}

Sensing and communication have been coexisting for decades. However, the traditional \textit{de facto} measure to avoid mutual interference between each other is the allocation of distinct spectrum bands for each of the two tasks. Despite being effective, such an approach is extremely spectrum-inefficient and can be even more problematic with the current scarcity in the spectrum resources \cite{surveyISAC}. Therefore, it has been critical to design spectrum-efficient techniques that can optimize the use of the available spectrum and resources while maintaining decent sensing and communication performance. In this context, integrated sensing and communication (ISAC) has emerged as a transformative paradigm, seamlessly unifying the traditionally separate domains of communication and sensing. Using shared spectral and hardware resources, ISAC systems not only enhance power and spectral efficiencies, but also enable a multitude of future sensing-enabled applications \cite{10812728}. An illustration of the interplay of ISAC use cases with emerging wireless technologies is presented in Fig. \ref{fig:ISAC_systems}. Although ISAC systems offer promising opportunities, their dual functionality introduces unique and complex security challenges. As these systems transmit sensitive data while simultaneously collecting environmental or object-related information, ensuring the security of both communication and sensing functionalities becomes paramount. 
Physical layer security (PLS), an approach that uses the inherent characteristics of wireless channels to protect data at the physical layer (PHY), has gained a significant amount of attention over the past few years. The key idea is to leverage PHY transmission techniques, stochastic channel coding, and the randomness in the wireless channel to establish information-theoretic secure transmission. Thus, PLS enhances data confidentiality without relying on computationally complex cryptographic schemes, making it energy-efficient and suitable for resource-constrained ISAC systems \cite{illi2023physical}. Furthermore, PLS serves as a theoretically secure barrier that fulfills the security needs for ISAC networks independently from the adversaries' computational power.

The scope of PLS in ISAC focuses on leveraging wireless channel characteristics and transmission techniques to secure communication and sensing functions. Jointly optimizing ISAC operations and PLS techniques' key parameters enhances system efficiency and privacy. PLS supports secure signal design, contributing to countering various types of attacks, such as sensing and data jamming, spoofing, and eavesdropping attacks. In addition to this, sensing-assisted secure ISAC communication has been recognized a fundamental use-case of ISAC to secure communication. In such an approach, sensing is established initially to locate the eavesdroppers (Eves) and to assist in estimating their channel state information (CSI), which is typically challenging to acquire in conventional communication systems due to the attackers' passive nature \cite{10227884}. A robust system design can enable continuous tracking of malicious nodes through sensing while communicating securely with legitimate users.

In this survey, we provide a comprehensive review of the state-of-the-art advancements in PLS techniques for ISAC systems focusing on three fundamental components of PLS: \textbf{confidentiality}, which ensures the protection of transmitted information from unauthorized access, and \textbf{covertness}, which aims to conceal the very existence of communication to enhance security against detection and interception, and \textbf{authentication}, which protects against unauthorized access and radar spoofing.


\subsection{Related Work}

The literature presents several studies and related surveys, tutorials, and brief reviews that explore security challenges in ISAC, either directly or partially related to PLS, such as one of the notable work by \textit{Furqan et al.} \cite{Furqam:OJCS:2021}, in which the authors discuss the security perspective of integrated wireless communication, sensing, and radio environment mapping (REM) in next-generation networks. The authors identify and classify the main communication and sensing security threats, such as eavesdropping, manipulation, and jamming, while suggesting some mitigation strategies, including low probability of intercept (LPI)-based transmission methods, PLS, and several PHY transmission techniques.
Nonetheless, despite the provided attacks' taxonomy and solutions, the work lacks (i) a formal introduction to ISAC and its types and (ii) mathematical foundations for ISAC systems analysis. Additionally, the work does not provide a comprehensive and system-wise review of existing schemes and/or techniques to realize secure ISAC networks, which is crucial for understanding the theoretical underpinnings and recent advancements in this domain.

Furthermore, \textit{Wei et al.} \cite{wei2022toward} provide a high-level review highlighting the security challenges for ISAC networks and reviews state-of-the-art security approaches, emphasizing the use of sensing capabilities to enhance PLS. The authors also propose low-cost secure ISAC architectures and outline open research areas, with an emphasis on robust, hardware-efficient designs suitable for low-cost ISAC devices. Nevertheless, a thorough literature review on the proposed secure ISAC schemes and a theoretical background for the analysis of ISAC networks' sensing and security performance were not covered.


Recently, our work \cite{illi2023physical} offers a comprehensive survey and tutorial on PLS techniques applied to the Internet of Things (IoT) networks, covering aspects such as authentication, confidentiality, and intrusion detection. Nonetheless, the aforementioned work dedicates only a small section to PLS techniques applied in ISAC systems. In addition, the authors of \cite{9524814} provide a holistic review of security threats in 6G networks as well as a taxonomy of solutions to mitigate them. Despite that, the work provides a limited discussion on the security of the sensing part in futuristic wireless networks, without presenting a detailed state-of-the-art on the proposed secrecy-enabling schemes in the context of ISAC systems. The work in \cite{9705498,9737357} presents detailed reviews of the various ISAC architectures and types, as well as formulated the fundamental theoretical limits of the different ISAC architectures. Nonetheless, it is worth emphasizing that there was a limited discussion on secure ISAC networks. Lastly, the survey work of Wei \textit{et al.} in \cite{10012421} provides a holistic review of the various waveform types and signal processing methods for ISAC networks. However, the security aspect of ISAC networks was neglected.

Despite the efforts provided in the aforementioned related surveys and tutorial work in ISAC-enabled networks, summarizing the various adopted ISAC architectures and schemes, a study and system-wise categorization of PLS-based schemes for ISAC networks is an apparent gap. 
Moreover, it is important to underscore the substantial body of recent literature published over the past two to three years, proposing novel physical layer security (PLS) schemes within the context of ISAC networks, which were not addressed in the work of \cite{Furqam:OJCS:2021}. Consequently, there is a critical need to (i) present a comprehensive review of these advanced PLS techniques and (ii) offer a system-level analysis along with a well-structured taxonomy of the various physical-layer methods that enable theoretically secure communications.

Motivated by the above, the aim of this paper is to provide a review of secure ISAC research, classifying the various PLS techniques employed in the field. In particular, this work reviews the various PLS schemes looking into confidentiality, covertness, and authentication, from both communication and sensing perspectives, catered by the utilization of cutting-edge technologies, such as multiple-input multiple-output (MIMO), reconfigurable intelligent surfaces (RIS), non-orthogonal multiple access (NOMA), and non-terrestrial communication (NTC). This classification aims to help readers easily navigate the existing literature and identify potential research gaps in the respective PHY techniques and network architectures considered in the secure ISAC designs. Furthermore, the current work also serves as a tutorial by providing a mathematical foundation for researchers interested in designing secure ISAC schemes. Additionally, while secrecy in ISAC networks has gained increasing attention over the past 5-6 years, the most recent related survey was published three years ago. Since then, numerous new studies have introduced diverse techniques and innovative approaches. Hence, an updated literature review is both timely and essential to synthesize recent advancements and guide future research in this rapidly evolving domain.
Table \ref{tab:comparison_related_works} provides a tabular comparison of the current work with the closely related work discussed above.  

\begin{table*}[htb!]
 \centering
 \caption{A Crisp Comparison of Closely Related Published Surveys and Tutorial Papers Concerning PLS and ISAC}
 \begin{tabular}
 {|p{0.8cm}|p{1.1cm}|p{1.6cm}|c|c|c|c|c|p{1.6cm}|c|p{1.2cm}|c|p{1.6cm}|}
 \hline
 \textbf{Work} & \textbf{Nature} & \textbf{Focus Area}   & \multicolumn{5}{c|}{\textbf{ Literature Review/Discussion on ISAC Systems}} &\textbf{Mathematical Background  (ISAC)}& \multicolumn{3}{c|} {\textbf{PLS Components}} & \textbf{Mathematical Background (PLS)}  \\
  &     &  & \textbf{MISO} &\textbf{MIMO}&\textbf{NOMA}&\textbf{RIS}&\textbf{NTC} & & \textbf{Conf.} &\textbf{Covertness} &\textbf{Auth.} &   \\
 \hline
 \cite{Furqam:OJCS:2021} (2021) & Perspective &  Security in wireless sensing, and  REM & - &-&- &$\checkmark$ &$\checkmark$&-&$\checkmark$ & $\checkmark$& $\checkmark$ &-\\
 \hline
 \cite{9524814} (2021)& Survey &Security and privacy for 6G &- &- &-&- &- &-& $\checkmark$& - & $\checkmark$&-\\
 \hline
 \cite{wei2022toward} (2022)& Overview &Securing ISAC transmission &- &- &-&- &- &-& $\checkmark$& -& $\checkmark$&- \\
 \hline
 \cite{9705498} (2022)&Survey \& Tutorial &Fundamental limits of ISAC &$\checkmark$& $\checkmark$&-&-& $\checkmark$&$\checkmark$ & -&-& -&-  \\ 
 \hline
 \cite{9737357} (2022)&Survey \& Tutorial &ISAC for 6G and beyond &$\checkmark$&$\checkmark$&-&$\checkmark$& -&$\checkmark$ & -&-&-&-  \\
 \hline
 \cite{illi2023physical} (2023)&Survey \& Tutorial &PLS in IoT &-& -&-& -&-&- & $\checkmark$&-&$\checkmark$ &$\checkmark$  \\
 \hline
 \cite{10012421} (2023)&Survey \& Tutorial &ISAC Signals &$\checkmark$& $\checkmark$&-& -&-&$\checkmark$ & -&-&-&-  \\
 \hline
 This work & Survey \& Tutorial & PLS in ISAC Systems &$\checkmark$ &$\checkmark$&$\checkmark$ &$\checkmark$&$\checkmark$ &$\checkmark$ &$\checkmark$ &$\checkmark$ &$\checkmark$&$\checkmark$ \\
 \hline
 \end{tabular}
 \label{tab:comparison_related_works}
 \end{table*}

\subsection{Contributions}

To the best of our knowledge, no survey currently exists that systematically examines and synthesizes the existing body of work on PLS approaches specifically designed for ISAC systems. To fill this gap and address this research need, our contributions in this paper are threefold:
\begin{itemize}
    \item First, we provide a foundational understanding of PLS, discussing three components of PLS (confidentiality, covertness, and authentication) and their mathematical background.
    \item Second, we present a detailed and system-oriented review of the state-of-the-art PLS techniques employed in ISAC. To this end, we offer a concise summary of notable work, systematically categorized, along with a thorough tabular comparison of these approaches to highlight their distinctive features and contributions.
    \item Finally, we explore emerging and innovative directions for future research in this field. In doing so, our goal is to inspire further advancements and foster the development of novel PLS strategies that effectively address the unique challenges posed by ISAC systems.
\end{itemize}

\subsection{Organization}
\begin{figure}[htb!]
     \centering
     \includegraphics[width=\linewidth]{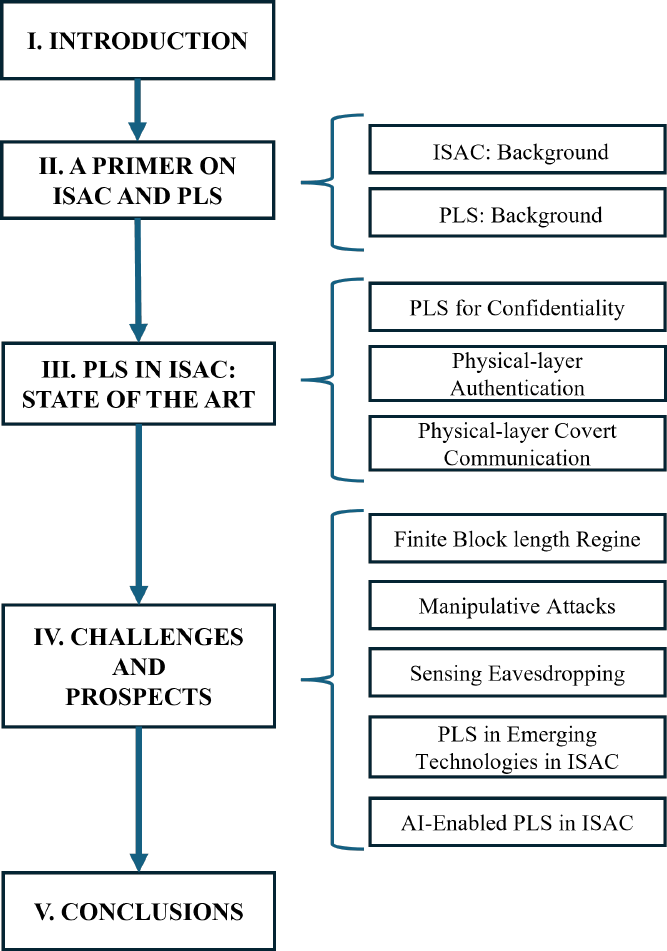}
     \caption{Organizational Structure of Our Work}
     \label{fig:org}
 \end{figure}

The rest of the paper is organized as follows: Section II provides background information on PLS and sensing, discussing various PLS and ISAC categories, along with the necessary mathematical foundations. Section III offers a comprehensive review, comparison, and analysis of the existing literature on PLS in diverse emerging ISAC systems. Section IV explores potential future research directions in the context of PLS in ISAC systems. Finally, Section V concludes the survey. A clear organizational structure or flow of the paper is presented in Fig. \ref{fig:org}.


\section{A Primer on  ISAC and PLS}


This section begins with a comprehensive overview of sensing principles and the diverse application domains ISAC systems. Subsequently, we delineate the core pillars of wireless security, highlighting the pivotal role of PLS in addressing their requirements through information-theoretically secure communication mechanisms. The final part of this section is devoted to establishing a rigorous theoretical foundation for evaluating the performance of ISAC-enabled networks from both sensing and security standpoints. This is achieved by employing key concepts from information theory, probability theory, and estimation theory. We anticipate that this foundational framework will serve as a valuable reference for researchers in the design and performance analysis of secure ISAC-based wireless systems.

\subsection{Integrated Sensing and Communication: Background}

The purpose of ISAC is to provide a power- and spectrum-efficient means for communicating and sensing using the same frequency/time resource and hardware modules. In general, there are two main techniques to establish ISAC networks, namely (i) radar-communication coexistence (RCC) and (ii) dual-function radar-communication (DFRC) \cite{surveymasouros}. 

In the former ISAC systems category, the radar and communication transceivers are separate but operate on the same frequency band. Therefore, the operation of RCC-based systems relies on designing smart interference management methods in order to allow both transceivers (radar and communication) to co-exist without harming each other with undesired interference. Within this ISAC category, several techniques have been proposed, such as

\begin{enumerate}
    \item \textbf{Opportunistic Spectrum Access}: Such a technique is an extension of the well-known cognitive radio networks, where the communication transceiver senses the spectrum use by the radar transceiver, considered as a primary user, in order to transmit its data signals.
    \item \textbf{Interference Channel Estimation}: Herein, the objective is to estimate the interference channel (radar-communication transceiver channel) in order to facilitate removing the interference effect of radar side lobes. Herein, pilot signaling by the radar transceiver is needed and several channel estimation techniques can be employed by the communication transceiver, such as the least-squares and minimum MSE estimators.
    \item \textbf{Robust pre-coding/Beamforming}: By exploiting the estimate of the interference channel, a robust beamforming design can be implemented at either the radar or communication transceivers to ensure a projection of the radar/communication signal into the null space of the radar/communication-communication/radar transceivers channel.
\end{enumerate}

On the other hand, DFRC enables the transmission of both sensing and communication signals by the same entity. Differently from the RCC category requiring a cooperation between the radar and communication entities, DFRC paves the way for a centralized processing at the same unit by exploiting the same signal for both aforementioned tasks. Herein, two general techniques can be employed within the ISAC-DFRC framework
\begin{itemize}
    \item \textbf{Temporal/Spectral Processing}: Information bits can be modulated onto radar pulses by using pulse interval modulation, or onto the slope of the chirp signals. A simpler way is to allocate different time slots for sensing and communication signals (i.e., time division).

    \item \textbf{Spatial Processing}: MIMO arrays in communication has been widely used in wireless communication network designs, which can enable signal beamforming as well as spatial multiplexing gains. Notably, MIMO application in radar allows for additional degrees of freedom in target estimation and enhances sensing performance. Thus, one straightforward use of MIMO arrays to establish DFRC-based ISAC systems is to convey information signals in the side lobes of the radiated beams, while exploiting the main lobes for sensing.
\end{itemize}

\subsubsection{Mathematical Background}

In the sequel to this section, we summarize several KPIs and metrics used for assessing the performance of secure ISAC networks. Mainly, metrics representing the secrecy and sensing performance are presented.

\paragraph{Target Detection}

As far as target detection is concerned, the objective is to exploit electromagnetic waves' reflection nature through objects to confirm the presence of a target at a given direction (i.e., AoA). Therefore, the DFRC transmitter aims at accurately beamforming towards the sensed targets, while transmitting information signal beams to the communication receivers. 

\begin{figure}
\centering
\begin{subfigure}
         \centering
         \includegraphics[scale=.5]{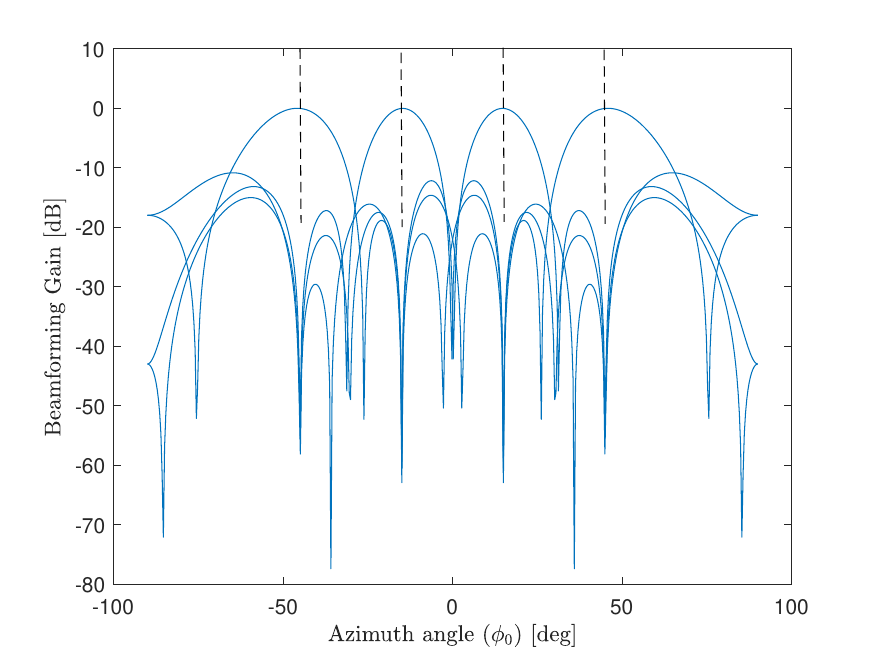}
         \caption{$N=8$ transmit antennas.}
         \label{fig6a}
     \end{subfigure}
\begin{subfigure}
         \centering
         \includegraphics[scale=.5]{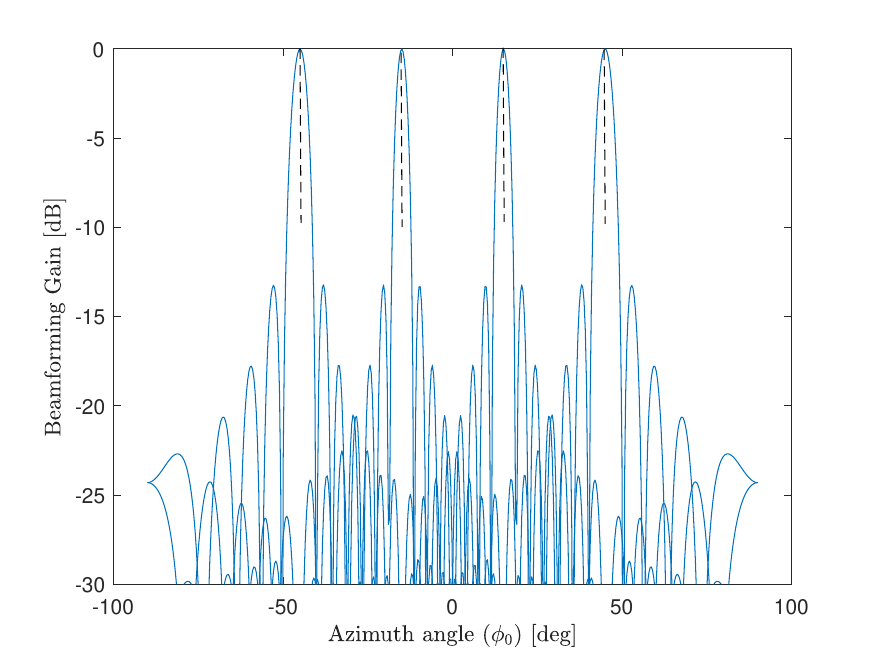}
         \caption{$N=32$ transmit antennas.}
         \label{fig6b}
     \end{subfigure}
\caption{Normalized beampattern gain plot. The figures illustrate the case of $L=3$ targets located at the directions $\theta=-\pi/4$, $-\pi/12$, and $\pi/12$, and $M=1$ communication user located at  $\theta=\pi/4$ (As indicated by dashed vertical lines). MRT is adopted as an ISAC beamformer.}

\label{fig6}
\end{figure}

\begin{itemize}
    \item \textbf{Normalized Beampattern Gain:} Such a metric measures the beam directivity of either the transmitted signal beams emanating from the transmit array towards the receivers and targets or the received beam gain from some particular directions. For a wireless ISAC network of $N_t$ transmit antennas, aiming to sense $L$ targets while communicating with $M$ communication users, such a measure can be expressed as 
\begin{align}
    P_b \left(\theta\right)&=  \mathbb{E}\left[\left|  \mathbf{h}_t \left(\theta\right)\mathbf{x} \right|^2 \right]
    \\ & = \mathbf{h}_t\left(\theta\right) \mathbf{R_x} \mathbf{h}_t^{H}\left(\theta\right),
\end{align}
where $\mathbf{h}_t\left(\theta\right)$ is the phased array response for the DFRC transmitter array with respect to a given looking direction $\theta$, which can be expressed for a uniform linear array as
\begin{equation}
    \mathbf{h}_t\left(\theta\right)= \left[ h_{t,1}\left(\theta\right),\ldots,h_{t,N_t}\left(\theta\right)\right],
\end{equation}
where
\begin{equation}
    h_{t,n}\left(\theta\right)=\exp\left( -j 2\pi \Delta (n-1) \sin\left(\theta \right)/ \lambda\right),n=1,\ldots,N_t,
\end{equation}
$\Delta$ is the transmit array's inter-element spacing, and $\lambda$ is the operating wavelength. The ISAC signal vector, beam-formed in transmission $\mathbf{x} \in \mathbb{C}^{N_t \times 1}$ is characterized by its covariance matrix $R_x \triangleq \mathbb{E}\left[\mathbf{x} \mathbf{x}^H\right]$ that defines the assigned power level for each transmitted signal and the correlation between the propagated signals. Such a covariance matrix can be designed in such a way to favoritize beamforming in particular directions of interest, e.g., the $L$ directions of the sensed target and $M$ ones of the communication users. By precoding a signal vector aimed at the $L$ targets and the single receiver as $\mathbf{x}=\mathbf{W}\mathbf{s}$, with $\mathbf{W} \in \mathbb{C}^{N_t \times (L+M)}$ is the precoding/beamforming matrix and $\mathbf{s}\in \mathbb{C}^{ (L+M) \times 1}$, one can elaborate several beams reaching the desired $L$ targets and $M$ users. For instance, maximal-ratio transmission (MRT) or zero-forcing (ZF) beamforming can be used as a choice for $\mathbf{W}$, which can offer spatial filtering by separating the various communication and sensing signals in the various angular directions of the various communication users and targets, i.e.,
\begin{equation}
    \left[\mathbf{W}^{\mathrm{(MRT)}}\right]_{:,i}=\frac{\mathbf{h}^H_{t} \left(\theta_i\right)}{\left\Vert \mathbf{h}_{t} \left(\theta_i\right)\right\Vert}, i=1,\ldots,M+L,
\end{equation}
\begin{equation}
    \left[\mathbf{W}^{\mathrm{(ZF)}}\right]_{:,i}=\frac{\left[\mathbf{H}^H_{t} \left( \mathbf{H}_{t} \mathbf{H}^H_{t}\right)^{-1}\right]_{:,i}}{\left\Vert \left[\mathbf{H}^H_{t} \left( \mathbf{H}_{t} \mathbf{H}^H_{t}\right)^{-1}\right]_{:,i}\right\Vert}, i=1,\ldots,M+L,
\end{equation}
where $ \left[\mathbf{X}\right]_{:,i}$ refers to the $i$th column of a matrix $\mathbf{X}$, $\left\Vert .\right\Vert$ indicates the Frobenius norm of a vector, $\mathbf{H}_t\triangleq \left[ \mathbf{h}_t \left(\theta_1 \right),\ldots,\mathbf{h}_t \left(\theta_{L+M} \right)\right]$, and $\theta_i$ is the angle of arrival (AoA) of the $i$th node (target or user) with respect to the DFRC transmitter. Fig. \ref{fig6} illustrates the normalized beampattern gain (divided by the maximal beampattern gain) for the case of $L=3$ targets and $M=1$ communication receiver. Notably, the increase of the transmit array's size enhances the beam directivity with a narrower half-power beamwidth and reduces the side lobes' power.

\item \textbf{Received Echo Signal Power:}
Signal beams emanating from the DFRC transmitter are designed to hit specific targets of interest. The radar receiver leverages the reflected echo signals from such targets in order to confirm their presence in such specific direction. Thus, it is crucial to assess the level of power collected by the receive radar array at the DFRC transceiver for a robust design of ISAC systems, considering the communication and sensing trade-off. The received power flux density (in $\mathrm{W}/\mathrm{m}^2$ at the $i$th target's reception plane can be expressed as\cite [Chapter 8, Eq. (8.70)]{emilbook}
\begin{equation}
    F_i= \frac{P_t P_b(\theta_i) }{4 \pi d_i^2},
\end{equation}
where $P_t$ is the DFRC transmitter's power, and $d_i$ is the distance between the DFRC transmitter and the $i$th target. The latter is characterized by an effective area called the radar cross section (RCS), denoted by $\sigma_{\mathrm{RCS}}$. Such a quantity, measured in $\mathrm{m}^2$ reflects how the sensed target reflects the incoming electromagnetic waves back towards the receiver. The RCS depends essentially on the object's physical properties and its location/orientation with respect to the transmitter. Consequently, the received echo signal's power back to the radar receiver can be expressed as 
\begin{equation}
    P_{r,i}= \frac{F_i \sigma_{\mathrm{RCS}} \left|\mathbf{u}_i \mathbf{h}_{r,i} \left( \theta_i\right) \right |^2\lambda^2}{(4 \pi d_i )^2},
   \label{rxechopower}
\end{equation}
where $\mathbf{u}_i \in \mathbb{C}^{1 \times N_r}$ is the radar receive filter for combining the $i$th target's echo signal, $N_r$ is the number of receive antennas, and $\mathbf{h}_{r,i} \left( \theta_i\right)\in \mathbb{C}^{N_r \times 1}$ channel response vector between the $i$th target and the radar receiver.

\item \textbf{Detection Probability:}
Target detection is a sensing identification problem whose objective is to confirm a target's presence at a particular direction. Therefore, it can be well modeled as a binary hypothesis testing problem, which can be expressed as
\begin{align}
    \mathcal{H}_{0,i} & : {y_i}[k]={n}[k] \notag \\
    \mathcal{H}_{1,i} & : {y_i}[k]=\sqrt{P_{r,i}} c_{\mathrm{RCS}} + {n}[k],
    \label{hypo}
\end{align}
where $\mathcal{H}_{0,i}$ indicates the $i$th target's absence at its pre-known direction, while $\mathcal{H}_{1,i}$ represents its presence, ${y}[k]$ refers to the $k$th time-domain sample of the received signal (after radar combining) from the $i$th target with $k=1,\ldots,K$ and $K$ is the number of collected echo samples for target detection, $c_{\mathrm{RCS}} \sim \mathcal{CN} \left( 0,1 \right)$ is a random variable representing the temporal fluctuations of the sensed object's RCS, and $n[k] \sim \mathcal{CN}(0,\sigma^2)$. Thus, it is observed from \eqref{rxechopower} and \eqref{hypo} that the $P_{r,i}$'s value is fundamental in correctly claiming either of the two hypotheses. By applying the Neyman-Pearson detection rule, one can formulate the target detection problem as \cite[Eq. (8.85)]{emilbook}
\begin{equation}
    \left|\mathbf{1}_K \mathbf{y}_i \right|^2 \gtrless \tau  ,
\end{equation}
with $\mathbf{1}_K$ is a row vector composed of the integer $1$ duplicated $K$ times $\mathbf{y}_i=\left[{y_i}[1],\ldots,{y_i}[K] \right]^T$, and $\tau$ is a decision threshold. 

In this context, two possible errors can be committed by the radar, namely the \textit{false alarm probability (FAP)} and the \textit{misdetection probability (MDP)}. While the former evaluates the probability of incorrectly claiming $\mathcal{H}_{1,i}$ while $\mathcal{H}_{0,i}$ takes place in reality, the latter evaluates the probability of misjudging a target's presence and claiming its absence, which can be formulated as
\begin{equation}
    P_{\mathrm{fa}}=\Pr\left[ \left.   \left|\mathbf{1}_K \mathbf{y}_i \right|^2 \geq \tau  \right|\mathcal{H}_{0,i}\right],
\end{equation}
and 
\begin{equation}
    P_{\mathrm{md}}=\Pr\left[ \left.   \left|\mathbf{1}_K \mathbf{y}_i \right|^2 < \tau  \right|\mathcal{H}_{1,i}\right].
\end{equation}
One can observe that the choice of the threshold $\tau$ affects both probabilities. Therefore, without loss of generality, a common approach is to derive the optimal threshold $\tau^{\ast}$ achieving a target FAP value, say $\delta$, and plug it into the MDP computation.
 
\end{itemize}

\paragraph{Target Localization}

Localization is another sensing application tasked with identifying the unknown location of a given target. Herein, in order to determine the unknown direction (i.e., AoA) of a given target, several beams are broadcasted at a set of angular directions, whereby the reflected echo is analyzed to estimate the AoA. Considering similar notations to the previous part on \textit{Target Detection}, the received average squared magnitude of the combined signal can be expressed as \cite[Eq. (8.4)]{emilbook}
\begin{align}
    Q\left(\mathbf{w}\right)&=\frac{1}{K} \sum_{k=1}^{K}\left|\mathbf{w}^H\mathbf{y}[k] \right|^2 \notag \\
    &=\mathbf{w}^H \left[\frac{1}{K} \sum_{k=1}^{K} \mathbf{y}[k] \mathbf{y}^H[k]\right] \mathbf{w}
    \label{rxpow}
\end{align}
where $\mathbf{y}[k]=\left[{y}_1[k],\ldots,{y}_{N_r}[k] \right]$ is the $k$th received echo signal vector sample at the $N_r$ receive antennas. Thus, we have 
\begin{equation}
    \mathbf{y}[k]=\frac{F \sigma_{\mathrm{RCS}} \lambda^2}{(4 \pi d )^2} \mathbf{h}_r\left( \theta_p\right) + \mathbf{n}[k],
\end{equation}
where the notations $F$, $d$, and $\mathbf{h}_r$ are similar to the target detection part presented earlier by only omitting the index $i$, while $\theta_p$ is the actual target's location (AoA). Below, two AoA-based localization techniques, suited for the scenarios of single- and multi-target localization are presented, based on maximizing the average squared magnitude of the received signal, computed in \eqref{rxpow}.

\begin{figure}
         \centering
         \includegraphics[scale=.5]{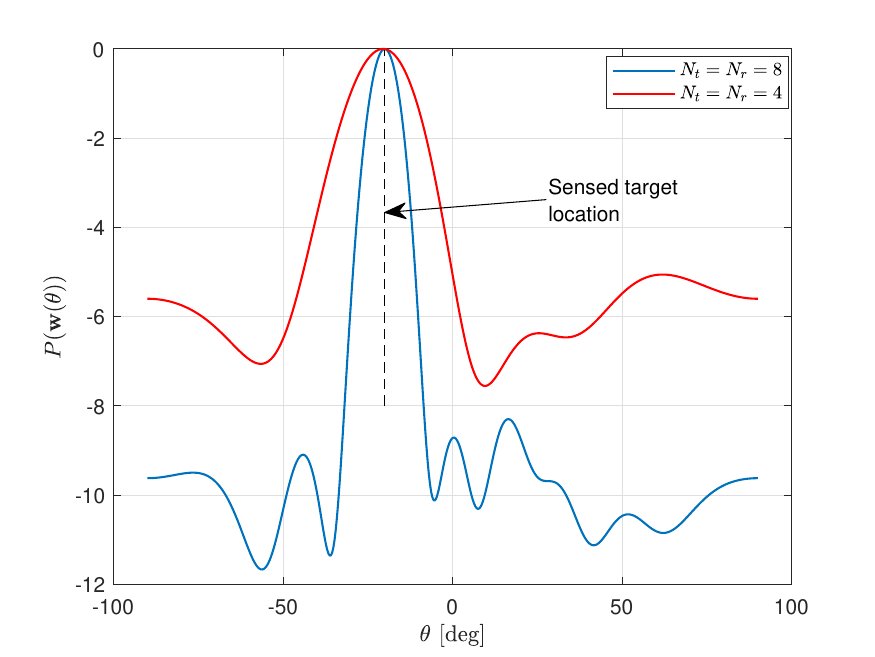}
         \caption{Spatial power spectrum obtained by evaluating \eqref{rxechopower} as a function of $\mathbf{w}(\theta)$ for varying $\theta$.}
         \label{fig7}
     \end{figure}

As far as a single target's case is concerned, non-parametric methods for the AoA estimation of a given target have been widely adopted \cite{emilbook}. Such methods aim at building a spatial spectrum similar to the frequency-domain spectrum adopted in spectral analysis techniques, e.g., Fourier transform. Herein, the receive combiner $\mathbf{w}$ is applied to the received signal with the objective of maximizing the power spectrum, given by \eqref{rxpow}, in the particular directions of interests (target's AoA). In \ref{rxpow}, we identify $\widehat{C} \triangleq \left[\frac{1}{K} \sum_{k=1}^{K} \mathbf{y}[k] \mathbf{y}^H[k]\right]$ as the sample estimate of the received signal's covariance matrix. In the absence of noise, observe from the first line of \ref{rxpow} that, according the Cauchy-Schwarz inequality, the power of the combined signal is maximized when $\mathbf{w}=\mathbf{h}_r\left( \theta_p\right)$. To this end, as $\theta_p$ is unknown, the combining vector is selected for a varying $\theta$ value computed from the tunable array phase response with a varying angle $\theta$, i.e., $\mathbf{w}\left(\theta\right)=\mathbf{h}_r\left(\theta \right)$. The latter value is plugged into \ref{rxpow} and evaluated at various $\theta$ values. Of note, the estimated AoA corresponds to the angular argument maximizing the combined signal power as
\begin{equation}
    \widehat{\theta}_p = \arg \max_{\theta \in \left[ -\pi/2, \pi/2\right]} P \left( \mathbf{w}\left( \theta\right) \right)
\end{equation}
Fig. \ref{fig7} shows the obtained power spectrum evaluated from \eqref{rxpow}. Herein, the evaluation is performed in terms of $\mathbf{w}(\theta)$ with a varying $\theta$ in $\left[ -\pi/2,\pi/2\right]$, whereas $K=50$. It can be observed that the obtained power spectrum exhibits a peak at $\theta=\theta_p=-20^{\circ}$, representing the true location of the target. The sensed target is illuminated by a zero-mean Gaussian-distributed omnidirectional signal vector of unit covariance matrix, covering the whole angular space. Observe also that the increase in the transmit and receive antenna arrays' size results in finer peaks, yielding a more accurate estimation of the target's AoA.

Thus, despite the conventional non-parametric method's acceptable performance in localizing a single target, its performance is limited in the case of $L>1$ targets. In particular, when the angular difference between the various targets is relatively small, the obtained power spectrum does not resolve efficiently the $L$ peaks from the various sensed targets' echoes. To this end, Capon beamforming is an alternative method that aims in localizing $L$ targets by means of optimizing the combining vector at the ISAC receiver (i) minimizing the received signal's variance and (ii) not impairing the reflected echo signal from the direction of interest (e.g., targets' AoAs). Thus, the overarching idea is to decrease the interference from other targets and directions by choosing the beamforming vector as \cite{capon}
\begin{equation}
    \widehat{\mathbf{w}} \left( \theta \right)= \frac{\widehat{C} ^{-1} \mathbf{h}_r \left( \theta\right) }{\mathbf{h}^H_r \left( \theta\right) \widehat{C} ^{-1} \mathbf{h}_r \left( \theta\right) }
\end{equation}
which results in
\begin{equation}
    P(\mathbf{w}(\theta))= \frac{1}{\mathbf{h}^H_r \left( \theta\right) \widehat{C} ^{-1} \mathbf{h}_r \left( \theta\right) }
    \label{rxpowcapon}
\end{equation}
\begin{figure}
         \centering
         \includegraphics[scale=.5]{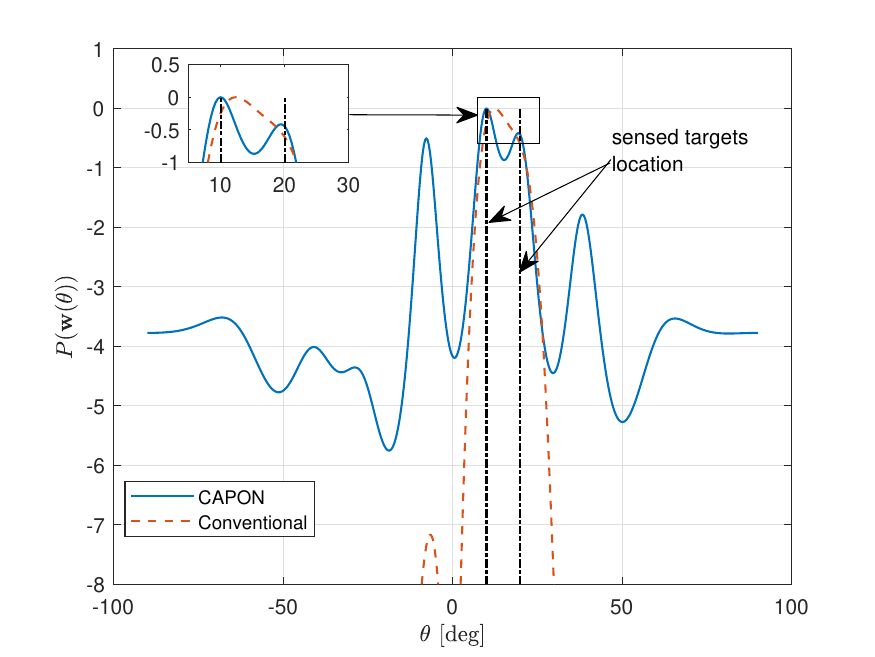}
         \caption{Spatial power spectrum obtained by evaluating \eqref{rxechopower} and \eqref{rxpowcapon} as a function of $\mathbf{w}(\theta)$ for varying $\theta$ with $L=2$ and $N_t=N_r=10$.}
         \label{fig8}
     \end{figure}

     Fig. \ref{fig8} shows the power spectrum obtained by the Capon method, evaluated from \eqref{rxpowcapon}, compared to the conventional beamforming method's power spectrum in \eqref{rxpow} for $L=2$, $K=20$, and $N_t=N_r=10$. Note that despite having two physical targets sensed, at directions of $10^{\circ}$ and $15^{\circ}$, the conventional matched-filtering beamforming fails in exhibiting two different peaks and manifests only a single peak at an AoA between both directions. On the other hand, the Capon method yields an angular separability of the two targets' peaks, yielding a more accurate estimation for $L>2$ targets' AoAs.

\subsection{Physical Layer Security: Background}
Every communication system is expected to meet five basic properties in the context of security, which are confidentiality, integrity, authentication, availability, and non-repudiation. With the growth of wireless communications and the proliferation of the IoT, securing data has become a pressing concern. Traditional cryptographic techniques, though widely effective, can be computationally expensive and may not be suitable for devices with limited processing power, such as sensors and IoT devices. Additionally, cryptographic methods often rely on assumptions about the computational infeasibility of certain mathematical problems (e.g., factorizing large numbers), which could be threatened by advancements in computing technologies, including quantum computing.

PLS provides an alternative by securing data based on the unique and dynamic characteristics of wireless communication environments. Since these characteristics vary between legitimate users and potential Eves, PLS can limit or prevent unauthorized access to the data at the PHY itself, making it an appealing option for scenarios with high privacy demands and resource-constrained devices.

\subsubsection{PLS for Confidentiality}
Physical layer confidentiality provides security to information in transit. In other words, this mechanism tackles eavesdropping attacks where Eve is in listening mode. The basis of this mechanism was first introduced by Wyner in his seminal work \cite{Wyner:BSTJ:1975}. The secrecy rate and the probability of leakage or secrecy outage (which depends on information-theoretic bounds) are the two main performance metrics for physical layer confidentiality. The secrecy rate can be enhanced through optimal resource allocation \cite{Waqas:WCNC:2016} and/or artificial noise (AN) generation \cite{ANPLS:WCSP:2016}, where noisy signals are generated in the null space of the legitimate users channel. Through similar procedures, secrecy outages can be minimized. Resources such as carrier power, carrier selection, relay selection, and user selection are exploited for physical layer confidentiality \cite{bloch2011physical}, \cite{Zhou:Book:2013}.

\subsubsection{{Physical-Layer Authentication}} Physical layer authentication (PLA) is a systematic procedure that verifies the legitimacy of the transmitter node based on the characteristics of the PHY, such as device fingerprints or features. PLA generally involves two steps: \textit{feature estimation} and \textit{testing} \cite{Bai:PLA:2020}. This mechanism requires a feature that is random in nature and different for distinct transmitters. The reported features can be mainly classified into medium/channel-based features and hardware-based features. Channel impulse response \cite{Ammar:VTC:2017}, channel frequency response \cite{CFR1:TWC:2008}, and received signal strength \cite{RSS:TPDS:2013} are examples of channel-based features. In-phase/quadrature (I/Q) imbalance \cite{IQ:ICC:2014} and carrier offsets \cite{Mahboob:VTC:2017} are examples of hardware features. Error probabilities (false alarm and miss detection) which also define receiver operating characteristics curves, Kullback-Leibler divergence \cite{CFR2:TWC:2012} and Jensen-Shannon divergence \cite{JSD:ICNNA:2017} are key performance metrics to evaluate PLA schemes.

\subsubsection{Physical-Layer Covert Communication}
Physical layer covert communication is a secure communication technique designed to ensure that a message is transmitted in such a way that its existence is hidden from unintended recipients or adversaries. Unlike traditional encryption, which secures the content of the message, covert communication focuses on concealing the communication itself. This is particularly important in scenarios where revealing the mere act of communication could be detrimental. The primary goal is to make the communication indistinguishable from background noise or normal network activity. Techniques often exploit PHY characteristics such as noise, channel variations, or interference. A communication channel specifically engineered to avoid detection by adversaries. These channels operate by embedding the communication signal within noise or exploiting unused system resources \cite{covertcomm}.


\subsubsection{Mathematical Background}
In the sequel, some essential mathematical foundations for analysis the secrecy performance of wireless networks are presented, including key performance metrics and core methods used to ensure the three main components of PLS. It is worth mentioning that such mathematical tools are readily applicable to ISAC networks.
\paragraph{PLS for Confidentiality}

The overarching idea of confidentiality-achieving PLS is to exploit stochastic channel encoding of the source message is order to convey a noisy observation to the Eve which cannot be decoded reliably (i.e., non-negligible decoding error probability at the Eve), while the legitimate receiver can reliably decode it, given its knowledge of the adopted channel coding scheme \cite{barros}. In order to ensure that the channel capacity of the legitimate link exceeds that of the eavesdropping link, the channel capacity is defined as the maximum data bits/symbols transmission rate such that the received message can be decoded with an arbitrarily small error probability, i.e., 
\begin{equation} 
    \underset{n \rightarrow \infty}{\lim} \underset{\Pr\left[ T \neq \hat{T}\right]}{\underbrace{P_d}} =0,
    \label{bereq}
\end{equation}
where $N$ is the codeword length corresponding to the adopted channel coding, $T$ and $\hat{T}$ are, respectively, the transmitted and decoded codewords, and $P_d$ is the codeword decoding error probability.

In order to define the SC, we consider a channel-coded communication between Alice and Bob (two legitimate parties) in the presence of an eavesdropper (Eve). Notably, the transmit rate depends essentially on the information message $M$ and the codeword's length $n$ as
\begin{equation}
    R=\frac{\mathbb{H}\left( M \right) }{n},
\end{equation}
where $\mathbb{H}\left( . \right)$ is the entropy operator. On the other hand, assuming Eve is aware of the used channel code, we define the equivocation at Eve as
\begin{equation}
    \Delta = \frac{\mathbb{H}\left( M \left| Z \right. \right)}{\mathbb{H}\left( M\right)},
\end{equation}
where $Z$ is the observed codeword at Eve. 

The secrecy capacity $C_S$ is defined as the maximum data rate fulfilling a reliable transmission between Alice and Bob while conserving, at least, a target equivocation rate $d$, such that $\Delta \geq d-\epsilon$ for small $\epsilon$ as \cite{barros}
\begin{equation}
    C_s= \underset{R}{\sup} \left(R,d \right).
\end{equation}

Reliability indicates that the probability of having the noisy observation of the intended receiver decoded in error is arbitrarily small, as detailed in (\ref{bereq}). Secrecy indicates a certain positive equivocation rate at the eavesdropper, such that the latter learns nothing about the data based on its noisy observation (i.e., the uncertainty about the message is unchanged after Eve receives its noisy observation). Per Wyner's weak secrecy model, it was shown that there exists a strictly positive secrecy capacity, allowing information-theoretically secure communication exclusively through channel coding when the quality of the main channel is better than that of the Eve channel \cite{DCR54, DCR55}. The secrecy capacity was shown to be the difference between the capacity of the main channel and the capacity of the Eve channel, considering discrete memoryless channels \cite{DCR54} and Gaussian channels subject to fading impairments \cite{DCR56}:
\begin{equation}
C_S = \text{max}\{C_C^{\text{(main)}} - C_C^{\text{(eve)}}, 0\}.
\end{equation}
For Gaussian channels, we have
\begin{equation}
C_C^{\text{(x)}}=\log_2\left( 1+\gamma^{\text{(x)}}\right),
\end{equation}
where ${\text{x}} \in \left\{{\text{main}},{\text{eve}} \right\}$, and $\gamma^{\text{(x)}}$ is the received signal-to-noise ratio (SNR) at Bob/Eve. Thus, an achievable secrecy rate is a data rate $R_S$ that is upper-bounded by the secrecy capacity as
\begin{equation}
R_S \in [0, C_S].
\end{equation}

The secrecy outage probability (SOP) is an extension of the secrecy capacity metric. It is defined as the probability that the secrecy capacity $C_S$ falls below a target secrecy rate $R_S$ for a given communication link in fading channels \cite{DCR57, DCR58}:
\begin{equation}
\text{SOP} = \text{Pr}[C_S < R_S].
\end{equation}

\paragraph{Physical-Layer Authentication}

In addition to data confidentiality, PLS techniques can be implemented for countering radar spoofing attacks, which are also known as radar manipulative attacks \cite{Furqam:OJCS:2021}. Herein, the attacker's objective is to inject a fake echo/sensing signal in the sensed environment in order to provoke an inaccurate estimation of one or several physical properties of the sensed environment or targets. For example, a properly injected echo signal can interfere with the legitimate one and lead to mis-evaluating the distance to a car in a vehicle-to-vehicle (V2V) wireless network, resulting in fatal consequences. Thus, the core interest of PLS here is to leverage the intrinsic PHY properties embedded in the received echo signal at the radar receiver to authenticate the legitimate echo signal and detect any sensing spoofing/jamming attacks. It should be noted that a parallel concept has been widely adopted at authenticating transmitting nodes, identifying their legitimacy, and detecting node spoofing attacks utilizing PHY attributes, which is known as physical-layer authentication (PLA). The overarching idea lies in building test statistics (TSs) based on the estimated PHY attribute from the received echo or information signal and an equivalent reference value estimate. The latter is evaluated or estimated based on an initial communication (echo sensing) signal received (reflected) from the legitimate user or target, and secured by upper-layer authentication mechanisms. Additionally, the reference value can be a purposely modified value in a transmitted challenge sensing signal by the ISAC transceiver, based on which a response is received and verified.
Thus, due to the similarity in applying PLA schemes for authenticating both communication and sensing signals, a generic and unifying framework can be presented accordingly.

As far as node or sensing signal authentication using PLA schemes is concerned, the ISAC receiver, acting as an authenticator, receives at a time instant $t$ an authentication request signal from an unknown source, from which a PHY feature value is estimated, i.e., $\widehat{W}^{\left(t \right)}_Z$, where $Z \in \left\{ S,I\right\}$ can be either the legitimate source $(S)$ or an intruder node $(I)$. The physical definition of $S$ refers to either a legitimate user or a legitimate sensed target, both of which can be spoofed by $I$. Thus, for authenticating the signal from an unknown source $Z$ at time $t$, the ISAC receiver extracts the considered PHY estimate from the received signal, i.e., $\widehat{W}^{\left(t \right)}_Z$, and utilizes the reference feature from the previous time instant observation or the feature value involved in the challenge signal, i.e., $(\hat{W}^{\left(t-1 \right)}_{Z})$. Accordingly, the source identification problem can be represented as a binary hypothesis testing one, with two potential hypotheses that can take place, namely:
\begin{itemize}
    \item $\mathcal{H}_0$: The sender is Alice (legitimate message), i.e., \begin{equation}
        \widehat{W}^{\left(t \right)}_Z=W^{\left(t \right)}_S+n_t,
    \end{equation}
    with $n_t$ denoting the receiver's estimation noise,
    \item $\mathcal{H}_1$: The sender is Eve (forged message), i.e.,
    \begin{equation}
        \widehat{W}^{\left(t \right)}_Z=W^{\left(t \right)}_I+n_t,
    \end{equation}
\end{itemize}

In order to decide on the actual hypothesis among the two potential ones, the likelihood ratio test (LRT) \cite{kaybook}
\begin{equation}
    \Psi = \log \frac{f_{\widehat{W}_Z^{(t)}\left| \mathcal{H}_1,\right.} \left( x\left| \mathcal{H}_1\right.\right)}{f_{\widehat{W}_Z^{(t)}\left| \mathcal{H}_0,\right.} \left( x\left| \mathcal{H}_0\right.\right)}
    \label{llrdef}
\end{equation}
where $f_{\widehat{W}_Z^{(t)}\left| \mathcal{H}_i,\right.} \left( x\left| \mathcal{H}_i\right.\right),i=0,1$ is the probability density function (PDF) of the feature estimate $\widehat{W}_{Z}^{(t)}$ conditioned on the occurrence of either hypotheses $\mathcal{H}_0$. For instance, the conditional PDF can be defined with a mean $\mu_0$ when conditioned on $\mathcal{H}_0$ and $\mu_1$ when conditioned on $\mathcal{H}_1$. Then, the ISAC receiver (authenticator) performs a simple comparison of the log-likelihood ratio $\Psi$ in \eqref{llrdef} with a predefined threshold ($\epsilon$) in order to infer on the legitimacy of the sending (reflecting) communication (sensing) source, i.e., deciding on $\mathcal{H}_0$ or $\mathcal{H}_1$ as
\begin{equation}
    \Psi \underset{\mathcal{H}_1}{\overset{\mathcal{H}_0}{\lessgtr}} \epsilon
    \label{HT}
\end{equation}

Due to the presence of noise, undesired propagation phenomena, and potential node and target mobility, signal reception is corrupted, which also impacts the PHY feature/parameter's estimation accuracy. Therefore, with such discrepancies in the estimated observations, two standard types of errors are evaluated, namely: 

\begin{enumerate}
    \item \textbf{False Alarm Probability (FAP)}: Also known as \textit{Type I Error}, which evaluates the frequency at which the null hypothesis $\mathcal{H}_0$ is incorrectly rejected after conducting the hypothesis testing (HT) in (\ref{HT}), which can be formulated as
    \begin{equation}
        P_{\mathrm{F}}=\Pr\left[\Psi \geq \epsilon \left\vert \mathcal{H}_{0}\right.  \right].
        \label{pfa}
    \end{equation}
The FAP represents the theoretical probability of incorrectly judging a legitimate communication or sensing signal as a spoofing one, whereby uts complementary is defined as the authentication probability, indicating the rate of correctly accepting a legitimate request or echo sensing signal.
     \item \textbf{Misdetection Probability (MDP)}: Such a probability defines the rate at which $\mathcal{H}_0$ is falsely claimed given that $\mathcal{H}_1$ occurred in reality, i.e., incorrectly accepting/authenticating a spoofing communication or sensing signal
    \begin{equation}
        P_{\mathrm{M}}=\Pr\left[\Psi < \epsilon \left\vert \mathcal{H}_{1}\right.  \right].
        \label{pmd}
    \end{equation}
  Similarly to the FAP, the complementary of the MDP is the detection probability, defined as the probability of correctly spotting a communication or sensing spoofing attack event.
\end{enumerate}

\paragraph{Physicla-Layer Covert Communication}

Despite the fact that the covert communication concept shares some similarities with the confidentiality-achieving PLS model, its main objective lies in achieving a transmission that is immune to being noticed by an eavesdropper or malicious warden device in the network. A typical covert communication system considers a legitimate transmitter Alice $(A)$ aiming to communicate privately with a legitimate receiver Bob $(B)$ without being noticed by a Warden node Willie $(W)$. The latter employs a radiometer to detect whether communication is taking place between $A$ and $B$, based on which this malicious entity decides to either perform eavesdropping or any other type of active attacks. Thus, $A$ targets minimizing the signal leakage to the Warden channel's and/or increasing the confusion level at $W$. Herein, $A$ relies on some sources of uncertainty, such as noise variance and channel uncertainty at $W$, stemming from the passive nature of $W$ and the fact that $W$ does not know anything about $A$. In addition, $A$ can use some artificial uncertainty sources, such as a random transmit power fluctuation. We denote the transmitted signal by $A$ as $x[k]$ of power $P_s$, and $k$ stands for the $k$th time-domain sample, whereby at each transmission round, $A$ decides to transmit her message with a probability $\pi_1$. The complementary event refers to the case when $A$ keeps silent, taking place with probability $\pi_0$. It is generally considered that $\pi_1=\pi_0=1/2$, representing the case of equiprobable transmission and silence events by $A$, maximizing the confusion at $W$. Consequently, the received signal at $W$ $y_W[k]=h_{AW}[k]x[k]+n_W[k]$, $h_{AW}[k]$ is the channel fading coefficient at the $k$th time slot between for the $A$-$W$ link, while $n_W[i]$ is the additive white Gaussian noise term at $W$ of variance $\sigma_W^2$. Due to the probabilistic transmission by $A$, $W$'s detection based on its received observation can be formulated as a binary hypothesis representation \cite{surveycovert}
\begin{equation}
    \begin{cases}
        &\mathcal{H}_0:y_W[k]=n_W[k]  \\
       &\mathcal{H}_1:y_W[k]=h_{AW}[k]x[k]+n_W[k]  \\
    \end{cases}
\end{equation}
where $\mathcal{H}_0$ represents the event when $A$ is silent while $\mathcal{H}_1$ defines the scenario when $A$ is transmitting. $W$ can decide on either hypothesis by measuring the average power of the received signal over $K$ time-domain samples as
\begin{equation}
   \underset{\triangleq P_r^{(W)}} {\underbrace{\frac{1}{K}\sum_{k=1}^{K} \left| y_{AW}[k]\right|^2}} \overset{\mathcal{H}_1}{\underset{\mathcal{H}_0}{\gtrless}} \tau.
   \label{decis}
\end{equation} 
It is usually assumed that (i)$W$ receives a sufficiently high number of samples ($K \rightarrow \infty$) in order to compute an accurate estimate of the average received signal power and (ii) a block fading assumption, rendering the channel coefficient constant over the received packet or $K$ samples, i.e., $h_{AW}[k]=h_{AW}, \forall k$. Thus, it yields $\triangleq P_r^{(W)}=P_s\left|h_{AW}\right|^2+\sigma_n^2$ In \eqref{decis}, $W$ decides $\mathcal{H}_1$ if the average received power $\triangleq P_r^{(W)}$ exceeds a certain threshold $\tau$, while $\mathcal{H}_0$ is decided otherwise. $A$ exploits $W$'s uncertainty about some of the parameters in the system, such as the noise variance or channel magnitude uncertainty, to confuse $W$'s decision. Accordingly, $W$'s hypothesis testing output can commit two types of errors, namely a false alarm and a misdetection, taking place with probabilities defined similarly to \eqref{pfa} and \eqref{pmd} as
\begin{equation}
        P_{\mathrm{F}}\left(\tau\right)=\Pr\left[  P_r^{(W)} \geq \tau \left\vert \mathcal{H}_{0}\right.  \right],
        \label{pfa2}
    \end{equation}
and
    \begin{equation}
        P_{\mathrm{M}}=\Pr\left[  P_r^{(W)} < \tau \left\vert \mathcal{H}_{1}\right.  \right],
        \label{pmd2}
    \end{equation}
where $P_{\mathrm{F}}\left(\tau\right)$ quantifies the rate at which $W$ mistakes Alice to be transmitting while the latter is silent, whereas $P_{\mathrm{M}}\left(\tau\right)$ evaluates the frequency by which $W$ misjudges $A$ to be transmitting. Accordingly, we define $P_{\mathrm{T}}\left(\tau\right)=P_{\mathrm{M}}\left(\tau\right)+P_{\mathrm{F}}\left(\tau\right)$ as the error detection probability. Intuitively, $A$ and $W$ have conflicting objectives on $P_{\mathrm{T}}$ where the former aims at increasing it as much as possible by virtue of the used uncertainty sources, while the latter aims at minimizing it. Observe that the choice of the threshold $\tau$ in \eqref{decis} is crucial as it defines the achievable FAP and MDP given by \eqref{pfa2} and \eqref{pmd2}. Generally, $A$ assumes that $W$ employs the optimal threshold $\tau^{\ast}$ guaranteeing a minimal error detection probability, i.e., $\tau^{\ast}=\arg \min_{\tau} P_{\mathrm{T}}\left(\tau\right)$. Then, $A$ aims at designing its transmission parameters, such as the power level selection, transmit beamforming, or RIS reflection pattern, to ensure $ \underset{\triangleq P_{\mathrm{T}}\left(\tau^{\ast}\right)}{\underbrace{P^{\ast}_{\mathrm{T}}}} \geq 1-\alpha$, where $\alpha$ is an arbitrarily small tolerance value.

\section{PLS in ISAC Systems: State-of-the-Art}
In this section, we categorize the state-of-the-art research on PLS in ISAC systems into three distinct subsections: PLS for confidentiality, PLS for covertness, and PLS for authentication. This classification provides a structured overview of existing work, facilitating a clearer understanding of security challenges and advancements in ISAC systems.

\subsection{PLS for Confidentiality}
We observed that PLS for confidentiality under the ISAC framework has been extensively studied in the literature over the past few years. Therefore, we categorize the discussion on PLS for confidentiality in ISAC systems based on different system configurations, including multiple-input single-output (MISO), MIMO, NOMA, RIS, and NTC. This system-wise classification provides a structured perspective on ongoing research, allowing for a clearer understanding of the key areas of focus and helping to identify research gaps in the field. Note that MISO, and MIMO, can also be explored in the context of NOMA, RIS, and NTC, therefore, their discussion is kept in relevant subsections only. 

\subsubsection{MISO-ISAC} 
\begin{figure}[htb!]
    \centering
    \includegraphics[width=\linewidth]{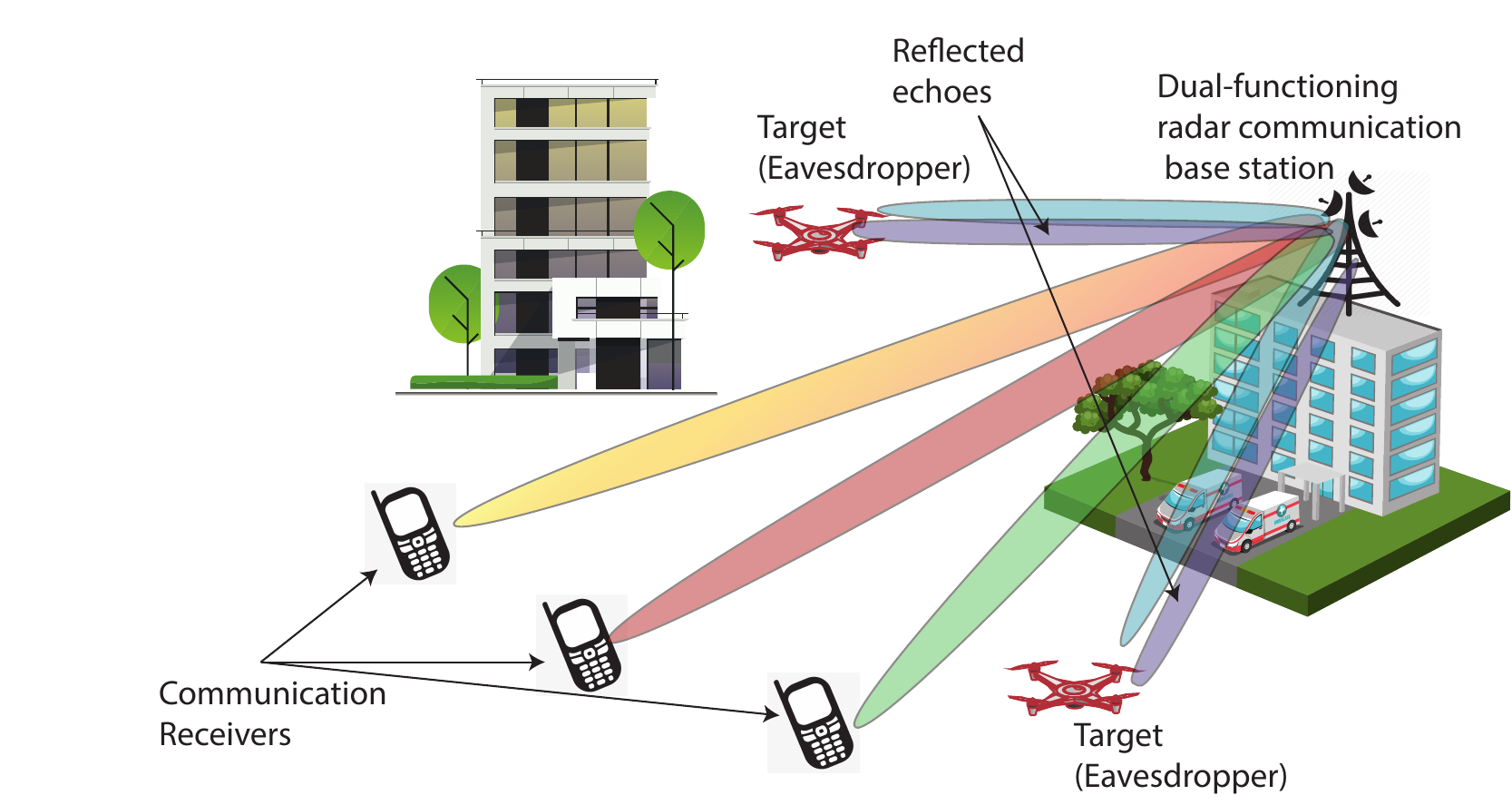}
    \caption{An illustration of a DFRC-based multi-user MISO ISAC system where a dual-function radar communication base station (BS) transmits signals to communication receivers while monitoring reflected echoes from targets, including potential Eves.}
    \label{fig:MISO_ISAC}
\end{figure}

In this subsection, we discuss work on PLS in ISAC systems that consider multiple antennas at the transmitter but single antennas at the communication users (CUs) in non-RIS, terrestrial, and OMA ISAC system configurations. We present an illustration of such a system with the presence of malicious targets/nodes in Fig. \ref{fig:MISO_ISAC}.

To the best of our knowledge, the first notable work on securing a MISO-ISAC system is reported by \textit{Su et al.} \cite{9199556}. They considered a DFRC system with a MISO setup where multiple CUs are considered equipped with a single antenna, a transmitter with multiple antennas, and a single target/Eve. AN is used to ensure secrecy by minimizing signal-to-interference-plus-noise ratio (SINR) at Eves while maintaining SINR for legitimate CUs. Variety of optimization techniques are developed for scenarios with perfect and imperfect CSI and varying levels of target location uncertainty. These include semi-definite relaxation (SDR) for converting non-convex problems into convex ones, fractional programming (FP) with Dinkelbach's transformation for efficiently solving fractional objective functions, and Robust Optimization to account for uncertainties in CSI and target location. The S-procedure is utilized to handle CSI errors, while the Lagrange dual function addresses statistical CSI uncertainty. Additionally, Eigenvalue decomposition and Gaussian randomization are used to approximate solutions from SDR, and the interior-point method efficiently solves the resulting semi-definite programs. The study also analyzes computational complexity and demonstrates the effectiveness of proposed algorithms through simulations, showing the trade-offs between radar and communication performance.

Following the previous work \cite{9199556}, \textit{Wen et al} \cite{wen2022joint} consider a single target and CU-assisted MISO setup for a DFRC system. They introduce a new secrecy measure known as secrecy estimation rate and formulate three secrecy rate maximization (SRM) problems, including SRM with and without AN and robust SRM, using secrecy rate and estimation rate as performance measures for communication and radar, respectively. The paper proves that the optimal beamformer for SRM can be computed in closed form. For the AN-aided SRM, the authors derive a closed-form solution for both the beamformer and the AN covariance matrix through alternating optimization (AO). Additionally, they consider the effects of imperfect CSI of the target, employing a moment-based random phase-error model for the direction of arrival. Simulation results validate the effectiveness and robustness of the proposed designs.

Building on previous work, \textit{Ren et al.} \cite{Ren:ICC:2022} consider a multi-antenna BS that communicates confidentially with a single-antenna CU while sensing multiple targets that may act as Eves. The BS transmits dedicated sensing signals that serve as AN to obscure Eves' efforts. The study aims to jointly optimize the transmit information and sensing beamforming to minimize the error between the actual and desired sensing beam-patterns, while meeting the minimum secrecy rate for the CU and adhering to power constraints. Although the problem is non-convex, the authors propose an algorithm leveraging SDR and a one-dimensional search to find a global solution. Additionally, two sub-optimal solutions based on zero-forcing and separate beamforming are introduced to reduce complexity. 
Further, the same authors in paper \cite{ren2023robust} extended their previous work to examine a downlink secure ISAC system where a multi-antenna  BS sends confidential messages to a CU while sensing potential eavesdropping targets. To protect communication, the BS uses dedicated sensing signals as AN to disrupt eavesdropping channels. Joint optimization of information and sensing beamforming is achieved to minimize beampattern errors while meeting secure communication constraints under two Eve CSI error models: bounded and Gaussian. For bounded errors, the study employs S-procedure, SDR, and 1D search, while Gaussian errors use Bernstein-type inequality and SDR. 

Next, \textit{Su et al.}  \cite{su2022secure} present a secure transmission method for DFRC systems in millimeter-wave networks. They consider a radar target as a potential Eve and employs constructive and destructive interference to enhance security. Directional modulation is used to leverage multi-user interference for secure communication. Constructive interference boosts legitimate CUs signals, while destructive interference disrupts Eve reception. The method involves FP and beamforming to optimize transmission under power and security constraints. 

This letter \cite{Liu:WCL:2022} studies secure multiple CUs downlink communication in ISAC systems and focuses on robust beamforming design under imperfect CSI for CUs and Eves, aiming to minimize beampattern error within secrecy outage and power constraints. The non-convex problem is addressed using SDR and Bernstein-type inequality. Results indicate the robust design closely matches perfect CSI performance and reveals trade-offs between sensing and secrecy.

Next, \textit{Dong et al.}  \cite{dong2023joint} propose a joint beamforming design for enhancing  PLS in DFRC systems. They focus on multiple CUs scenarios and show that MIMO radar waveforms can also function as AN to disrupt eavesdropping channels while ensuring effective target detection and communication. The proposed joint optimization of radar and communication beamforming is formulated as a non-convex problem aimed at balancing radar beam patterns, communication quality of service (QoS), and PLS. The paper employs SDR and a reduced-complexity algorithm to address the non-convexity while ensuring global optimization.

Further, \textit{Chu et al.}   \cite{Chu:TVT:2023} addresses PLS in a different ISAC system that shares a spectrum between multiple CUs communication and colocated MIMO radar (also known as \textbf{RCC}\footnote{RCC focuses on enabling radar and communication systems to share the same spectrum without compromising their independent functionalities. It emphasizes managing interference between the systems operating in overlapping frequency bands}). By exploiting radar signal interference, joint beamforming is optimized to minimize the maximum SINR at Eves while meeting communication quality, radar detection, and power constraints. An efficient algorithm based on the block coordinate descent (BCD), FP, and SDR methods is
developed to solve the resulting non-convex optimization
problem. When Eves' CSI is unknown, an AN-aided beamforming scheme uses residual power to disrupt eavesdropping while maintaining legitimate transmission and radar performance.

On the other hand, \textit{Yang et al.} in \cite{Yang:WCNC:2024} study secure resource allocation in a downlink ISAC system using \textbf{semantic communication}\footnote{Semantic communication is an emerging paradigm in the field of wireless communication that goes beyond traditional bit-level accuracy by focusing on the meaning, context, and intent behind the transmitted information} with multiple CUs and Eves. The BS uses beamforming to transmit sensing and communication signals, with sensing signals serving as AN for enhanced security. To boost security at the semantic level, semantic information is extracted and shared with users via a knowledge base stored beforehand. The objective is to maximize users' sum semantic secrecy rate while ensuring minimum quality of service and sensing performance. An iterative algorithm based on AO is proposed, showing superior results in secure semantic communication and target detection.
 
Next, this paper \cite{hou2024optimal} examines a secure ISAC system where a multi-antenna BS communicates with a single-antenna CU while sensing the location of a potential Eve target through reflected signals. The \textit{target's location is random, but its distribution is known}. The study derives the posterior Cramer-Rao bound (PCRB)\footnote{Unlike the classical Cramer-Rao Bound (CRB), which relies only on the likelihood function, the PCRB incorporates prior information about the parameter being estimated} for mean-squared error (MSE) in sensing, providing insights into the need for "probability-dependent power focusing" in beamforming. The optimization problem aims to maximize the worst-case (among all the possible locations for Eve) secrecy rate while maintaining a threshold on PCRB. This non-convex problem is addressed using a two-stage approach, transforming it into a convex form with SDR. 

\textit{Xu et al. }  \cite{Xu:IoTJ:2024} explore ISAC systems that support both public and confidential transmissions while simultaneously tracking targets. The focus is on an ISAC BS that manages these diverse services and aims to minimize the discrepancy between actual and desired beampatterns, adhering to constraints on public message rates and confidential message secrecy rates.
For time-invariant channels, where CSI estimation errors are minimal, the authors propose a SCA algorithm to jointly design the transmit covariance matrices and the AN covariance matrix. Additionally, they introduce a low-complexity two-stage algorithm for practical implementation. The study extends to time-varying channels, accommodating larger estimation errors in CSI.

On the other hand, \textit{Ren et al.}  \cite{ren2024secure} explore a secure \textbf{cell-free}\footnote{The concept of cell-free networks is a shift away from the traditional cellular architecture where each user is served by a single BS within a predefined cell.} ISAC system where multiple transmitters collaboratively send confidential information to CUs and conduct target detection while addressing threats from both communication and sensing Eves. The joint transmit beamforming is optimized to maximize detection probability for sensing targets, while meeting SINR requirements for information confidentiality and SNR limits to protect sensing privacy. This complex non-convex problem is solved using SDR to achieve the globally optimal solution. Additional beamforming designs based on sensing SNR maximization and coordinated beamforming are also proposed, with simulations highlighting the effectiveness of the primary design.

Following the previous work \cite{ren2024secure}, \textit{Ali} \cite{nasir2024joint} highlights the complexity of using SDR in previous work and investigates a secure cell-free ISAC system where distributed access points collaboratively serve CUs and sense a target while countering Eves. He proposes a joint optimization of communication and sensing beamforming to maximize both CUs secrecy rates and target sensing SNR. The non-convex problem is tackled using an iterative optimization algorithm based on quadratic programming. Simulations confirm the algorithm's superior sensing SNR and secrecy rate performance, particularly when Eves are close to users, and show that it achieves a high secrecy rate with reduced computational complexity compared to existing methods.

On the other hand, in another notable work \cite{Bazi:TIFS:2024}, \textit{Bazzi et al.} introduce a secure \textbf{full-duplex (FD)-ISAC}\footnote{FD-ISAC refers to a system that simultaneously performs sensing and communication in both directions (transmitting and receiving) over the same frequency band and at the same time.} system where a DFRC BS communicates with users while sensing Eves aiming to intercept uplink (UL) and downlink (DL) data.  They introduce an optimization framework that leverages AN to enhance UL and DL secrecy rates while optimizing the integrated sidelobe-to-mainlobe ratio (ISMR) for radar performance. A novel SCA algorithm iteratively solves the optimization problem, ensuring power efficiency and robust system performance.
 
Very recently, \textit{Su et al.} in a preprint \cite{su2024secure} introduce a signaling design for secure ISAC systems using a dual-functional MIMO BS that simultaneously communicates with multiple CUs and detects potential Eves. The design aims to minimize the \textbf{Bayesian Cramer-Rao bound (BCRB)}\footnote{It provides a fundamental performance limit for parameter estimation in Bayesian frameworks, where the parameters to be estimated are modeled as random variables with known prior distributions.} while ensuring QoS through constructive interference, thereby degrading the Eves' ability to decode. A tailored SCA method is employed to solve the non-convex optimization problem. The proposed scheme outperforms traditional block-level precoding techniques, providing improved PLS and sensing accuracy.

Finally, we provide a tabular summary and comparison of the existing state of the art on securing MISO ISAC systems via PLS in Table \ref{tab:comparison_isac}. Note that DFRC and ISAC terms are used interchangeably, depending on the terminology adopted in corresponding papers.

\begin{table*}[htb!]
\centering
\caption{Tabular Summary \& Comparison of PLS-based Secure ISAC MISO Systems}
\begin{tabular}{|p{3cm}|p{4cm}|p{3cm}|p{6cm}|}
\hline
\textbf{Reference} & \textbf{System Setup} & \textbf{Optimization Techniques} & \textbf{Features}  \\ \hline
Su et al. (2020)  \cite{9199556} & DFRC, Multiple CUs, Single Target as Eve & SDR, FP, Robust Optimization, S-Procedure &  Secrecy rate maximization through beamforming under channel uncertainties \\ \hline
Wen et al. (2022) \cite{wen2022joint}& DFRC, Single Target as Eve, and a Single CU & Closed-form solutions, AO, Random Phase-Error Model & Joint optimization of radar SINR and secrecy rate considering phase errors \\ \hline
Ren et al. (2022) \cite{Ren:ICC:2022} & ISAC, Multi-Antenna BS, Multiple Targets as Eves, Single CU & SDR, One-dimensional search, Zero-forcing &Secrecy rate enhancement through beamforming and zero-forcing at the BS \\ \hline
Ren et al. (2023) \cite{ren2023robust} & ISAC, Multi-Antenna BS, Multiple Targets as Eves & S-Procedure, SDR, Bernstein-type inequality &Secrecy performance optimization under bounded CSI errors \\ \hline
Su et al. (2022) \cite{su2022secure} & DFRC, Millimeter-wave CU, Single Target as Eve & FP, Beamforming & Secure beamforming design for mmWave DFRC with secrecy rate maximization \\ \hline
Liu et al. (2022) \cite{Liu:WCL:2022} & ISAC,  Multiple CUs, Multiple Targets as Eves & SDR, Bernstein-type inequality & Robust beamforming for joint communication and secure sensing with multiple eavesdroppers \\ \hline
Dong et al. (2023)   \cite{dong2023joint} & DFRC, MIMO Radar, Multiple CUs and Targets as Eves  & SDR, Reduced-complexity algorithm &Joint beamforming and artificial noise design for secrecy rate optimization \\ \hline
Chu et al. (2023)  \cite{Chu:TVT:2023}& RCC, Colocated MIMO Radar,  Multiple CUs and Eves, a single target & BCD, FP, SDR & Secrecy rate maximization through joint communication and radar waveform design \\ \hline
Yang et al. (2024) \cite{Yang:WCNC:2024} & ISAC, Semantic Communication, Multiple CUs and Targets as Eves & AO &Secure semantic communication and beamforming design for ISAC  \\ \hline
Hou et al. (2024) \cite{hou2024optimal} & ISAC, Multi-Antenna BS, Single CU, Single target as Eve (Prior location information is known) & SDR, Two-stage convex transformation & Location-aware secure beamforming for optimized secrecy and sensing performance  \\ \hline
Xu et al. (2024) \cite{Xu:IoTJ:2024} & ISAC, Public/Confidential Transmissions to UAVs treated as high-security clearance UAV as legitimate while low-security clearance UAVs as potential Eves & SCA, Two-stage algorithm & Secure transmission and sensing beamforming for heterogeneous UAV security levels \\ \hline
Ren et al. (2024) \cite{ren2024secure} & Cell-Free ISAC, Multiple ISAC Transmitters and Sensing Receivers (aka multi-static sensing), Multiple CUs and Information and Sensing Eves & SDR & Distributed beamforming design for joint communication and sensing secrecy enhancement \\ \hline
Ali (2024) \cite{nasir2024joint} & Cell-Free ISAC, Multiple ISAC Transmitters and Sensing Receivers, Single Target and Multiple Information Eve & Iterative quadratic programming & Joint optimization of secure information transmission and sensing in cell-free ISAC\\ \hline
Bazzi et al. (2024) \cite{Bazi:TIFS:2024} & FD ISAC, Multiple UL and DL CUs with Multiple Target as Eves  & SCA & Full-duplex joint uplink and downlink secure ISAC beamforming optimization \\ \hline
Su et al. (2024) \cite{su2024secure} & ISAC, MIMO BS, Multiple CUs and target as Eves & SCA & Secrecy rate optimization through joint transmit beamforming and sensing design \\ \hline
\end{tabular}
\label{tab:comparison_isac}
\end{table*}

\subsubsection{MIMO-ISAC}
In this subsection, we discuss work that address PLS in ISAC environments in non-RIS, terrestrial, and OMA systems where both the transmitter and CUs are considered equipped with multiple antennas.  MIMO is considered a promising technique to enhance the performance of emerging technologies in 6G networks \cite{9971740}.

To the best of our knowledge, the first notable work in this direction is reported by \textit{Deligiannis et al.}  \cite{deligiannis2018secrecy}. They explore transmit beampattern optimization for MIMO radar systems that must detect a multi-antenna target and communicate securely with a multi-antenna legitimate CU while protecting against an eavesdropping target. Three optimization problems are addressed: maximizing target return SINR, maximizing secrecy rate, and minimizing transmit power. Since the secrecy rate function is non-convex, \textit{Taylor series approximations} and iterative algorithms are used to transform the problems into convex ones. They design two transmit covariance matrices to balance target detection and secure communication, with simulations validating the approach.

Next, building on the previous work \cite{deligiannis2018secrecy}, \textit{chalise et al.}  \cite{chalise2018performance} present a unified system for \textbf{passive radar}\footnote{A passive radar is a type of radar system that does not actively transmit its own signal. Instead, it relies on existing signals from other sources, such as radio, television, cellular networks, or satellite communications, to detect and track objects. Passive radar systems exploit the reflections of these ambient signals from objects of interest.} and communication, addressing the risk of eavesdropping by a passive radar receiver. The objective is to maximize the SINR at the radar receiver while maintaining a minimum secrecy rate for the communication system. Both disjoint (non-overlapping) and overlapping transmission scenarios are considered, with non-convex optimization problems addressed using AO techniques. The paper employs semi-definite programming and semi-analytical methods for the disjoint case and an SDR approach for the overlapping case. Performance comparisons highlight the benefits of the overlapping approach and semi-analytical techniques.


\subsubsection{NOMA-ISAC}

\begin{figure*}
    \centering
    \includegraphics[width=0.8\linewidth]{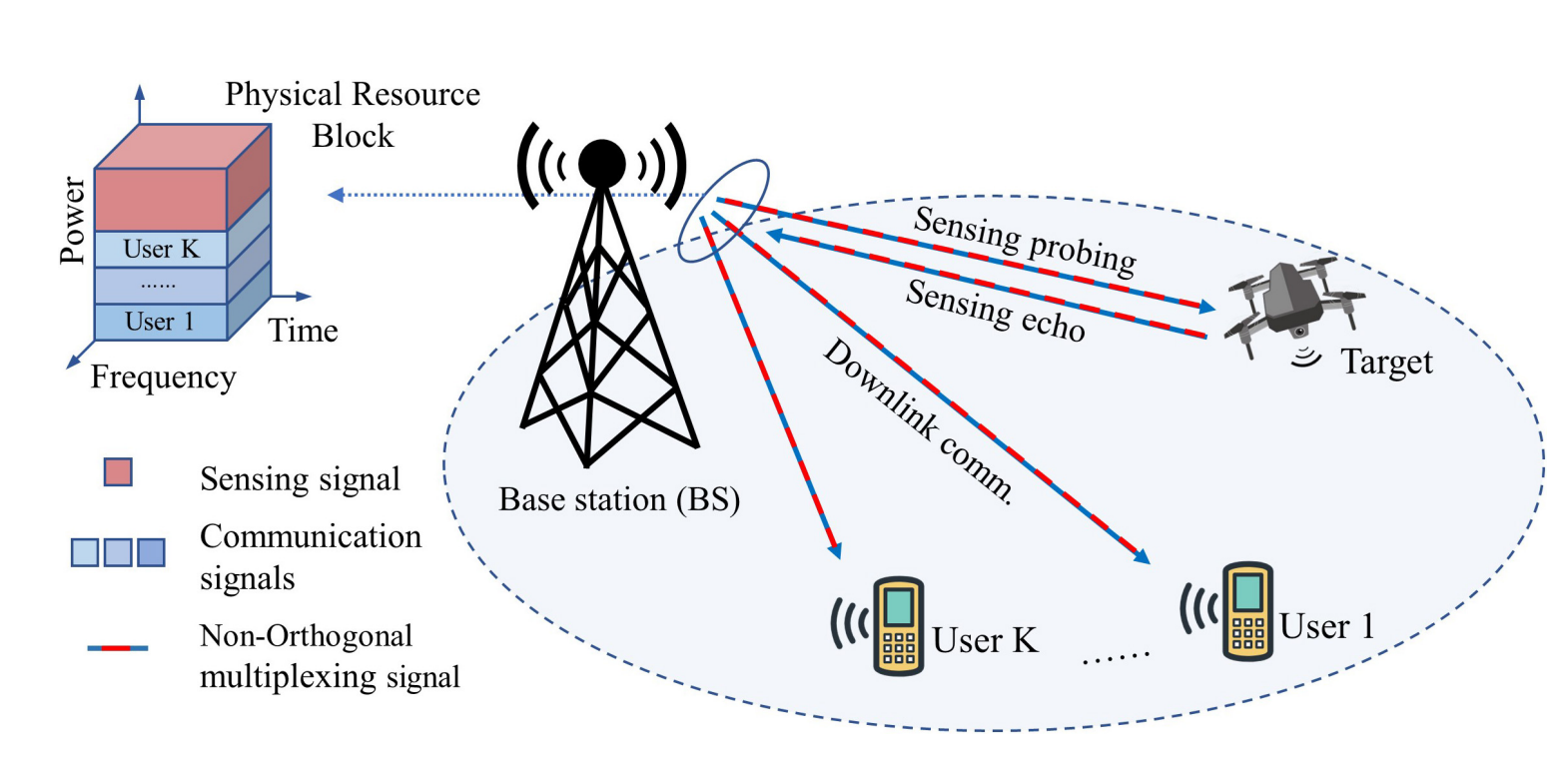}
    \caption{An illustration of ISAC system utilizing NOMA, where a BS transmits communication signals to multiple users and sensing signals to a target, achieving joint downlink communication and sensing via resource sharing in the time-frequency-power domain \cite{10237262}.}
    \label{fig:NOMA}
\end{figure*}
NOMA is an advanced multiple-access technique used in wireless communication systems to enhance spectral efficiency and user connectivity \cite{mu2022noma}. Unlike traditional orthogonal methods like TDMA or FDMA, NOMA allows multiple users to share the same time and frequency resources by employing power-domain or code-domain multiplexing as illustrated in Fig. \ref{fig:NOMA}. In power-domain NOMA, users are assigned different power levels based on their channel conditions, enabling simultaneous transmission. Successive interference cancellation (SIC) is a key receiver technique used to decode signals in NOMA, separating higher-power signals from lower-power ones. This enables better utilization of resources and supports a greater number of users compared to orthogonal methods. NOMA is especially promising for 5G and beyond, as it addresses the growing demand for massive connectivity in applications like IoT and smart cities. However, challenges like inter-user interference, SIC complexity, and security need to be addressed for widespread deployment \cite{8352619}.

In the context of PLS in NOMA-aided ISAC, to the best of our knowledge, \textit{Yang et al.} report the first study  \cite{yang2022secure}, which presents a secure pre-coding optimization for NOMA-aided ISAC systems. The proposed scheme aims to maximize the sum secrecy rate for multiple CUs while ensuring the sensing performance. AN has been introduced to enhance security, particularly against Eves, which target weak-channel NOMA users. A non-convex joint pre-coding optimization problem is formulated and solved using SCA, Taylor's approximation, and second-order cone programming (SOCP).

Building on the previous work, \textit{Luo et al.}  \cite{9854898} ensure secure transmission by concealing confidential user information within radar beams. Legitimate CUs remove interference using SIC, while Eves face interference due to the decoding order. The problem is formulated as a non-convex optimization task, which is decomposed into two sub-problems solved iteratively. Numerical results demonstrate the method's effectiveness in enhancing security and performance. 

Next, \textit{Liu et al.}  \cite{liu2024secure} claim that traditional methods fail when legitimate CUs are at close angles, and perfect CSI and SIC assumptions are unrealistic. The authors propose user clustering and beamforming to suppress inter-cluster interference and enhance signal decoding. They optimize power allocation using AO, SDR, and Iterative Rank Minimization (IRM) to solve a non-convex optimization problem. AN is employed for sensing and security, ensuring eavesdropping risks are mitigated. Simulation results show the proposed method outperforms traditional approaches in sensing and communication performance and is robust against channel imperfections and imperfect SIC, which are critical in practical scenarios.

Recently, following the previous work, \textit{Zhang et al.}  \cite{zhang2024secure} assume imperfect CSI and uncertainties regarding the angles of malicious targets and propose a secure beamforming scheme. In this context, multicast signals are used to convey group-oriented information to cooperative targets while also disrupting Eves and ensuring sensing capabilities. A robust beamforming optimization problem is formulated to address imperfect CSI and uncertainties regarding the angles of malicious targets. The proposed penalty-based iterative algorithm transforms the non-convex problem into a convex one using SDR. 

Finally, we provide a tabular summary and comparison of the existing state of the art on securing MIMO-ISAC and NOMA-ISAC systems using PLS in Table \ref{tab:comparison_mimo_noma_isac}.

\begin{table*}[htb!]
\centering
\caption{Tabular Summary \& Comparison of PLS-based Secure ISAC MIMO and NOMA Systems}
\begin{tabular}{|p{3cm}|p{4cm}|p{3cm}|p{6cm}|}
\hline
\textbf{References} & \textbf{System Setup} & \textbf{Optimization Techniques} & \textbf{Features} \\ \hline
Deligiannis et al. (2018) \cite{deligiannis2018secrecy} & MIMO Radar, Multi-Antenna Target as Eve, Multi-Antenna CU & Taylor series approximations, Iterative algorithms & Secrecy rate optimization through transmit covariance matrices of information and distortion signals\\ \hline
Chalise et al. (2018) \cite{chalise2018performance}& Passive Radar and Communication, Single Radar Receiver as Eve, Single CU and single Target & SDR, AO, Semi-analytical methods & Radar SINR optimization through waveform design  \\ \hline
Yang et al. (2022) \cite{yang2022secure} & NOMA-aided ISAC, Multiple NOMA CUs with Single Antenna Target as Eve & SCA, Taylor's approximation, SOCP & Secrecy rate maximization through joint beamforming and power allocation design \\ \hline
Luo et al. (2022) \cite{9854898} & NOMA-aided DFRC, Multiple NOMA CUs, Single Target, Single Eve & Iterative sub-problems, Non-convex decomposition & Secrecy rate enhancement through joint transmit beamforming and AN generation \\ \hline
Liu et al. (2024) \cite{liu2024secure} & NOMA-aided ISAC, Single Antenna NOMA CUs with Clustering, Single Antenna Multiple Targets as Eves  & AO, SDR, IRM & Secure and efficient joint clustering and beamforming design for maximized secrecy rate \\ \hline
Zhang et al. (2024) \cite{zhang2024secure} & NOMA-aided ISAC with Multicast Signals, Single CU and Single Radar Centric-User and Single Target as Eve & Penalty-based iterative algorithm, SDR & Secrecy rate optimization through multicast beamforming and power control  \\ \hline
\end{tabular}
\label{tab:comparison_mimo_noma_isac}
\end{table*}

\subsubsection{RIS-assisted ISAC}
\begin{figure}[htb!]
    \centering
    \includegraphics[width=\linewidth]{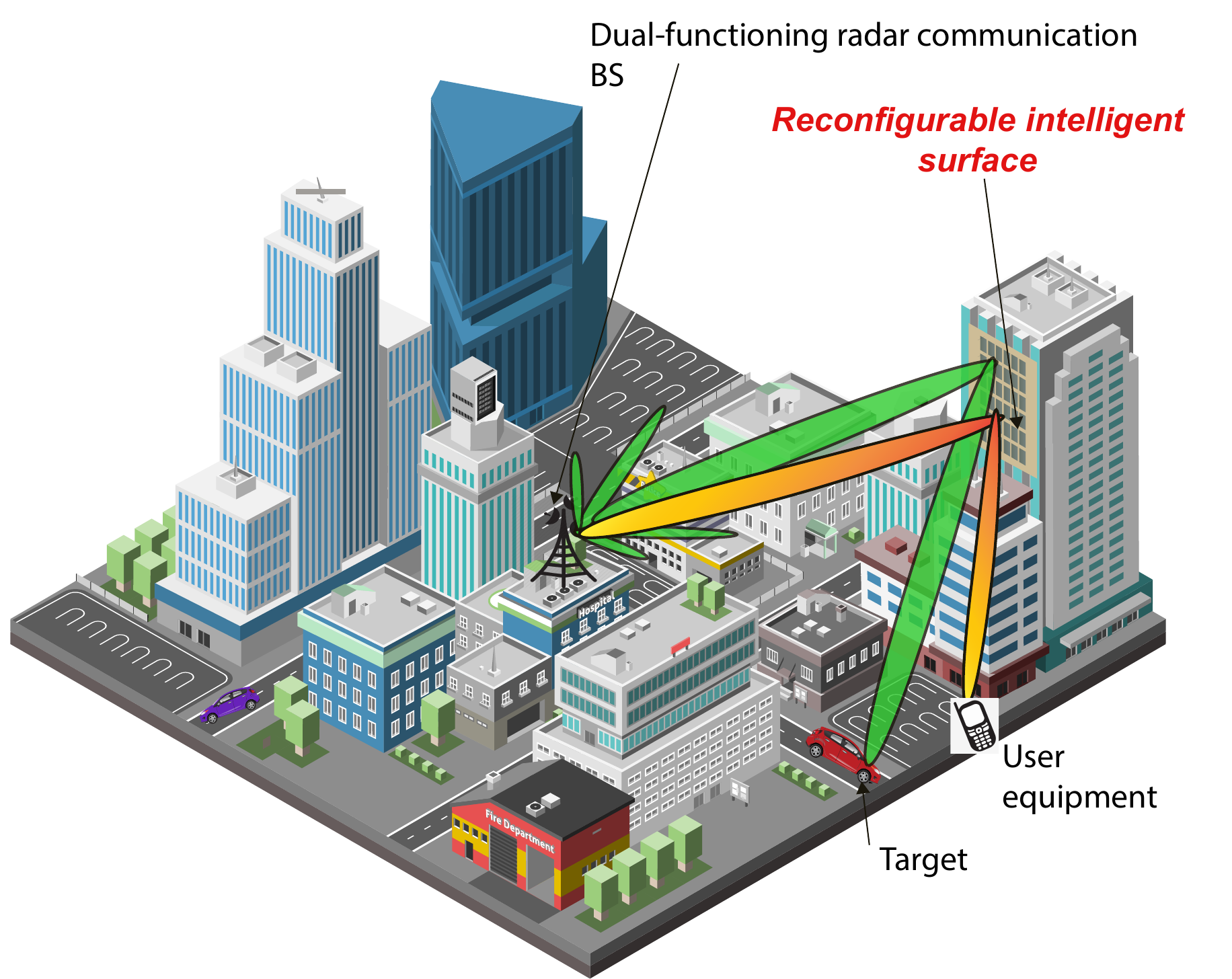}
    \caption{An illustration of a RIS-assisted ISAC network showcasing a RIS panel mounted on a building. The RIS panel simultaneously serves a CU by optimizing wireless signals and detects a target in the environment, highlighting its dual functionality of communication and sensing.}
    \label{fig:RIS-ISAC}
\end{figure}
RIS-assisted ISAC represents a novel and promising technology that combines the advantages of RIS with ISAC to enhance the performance of both communication and sensing systems. RIS is an emerging technology that can intelligently control the propagation environment by manipulating electromagnetic waves through a surface composed of numerous low-cost, passive, and tunable elements. When RIS is integrated with ISAC, it offers several benefits, such as improved signal quality, enhanced sensing capabilities, and more efficient use of the radio spectrum \cite{yu2023active}. An illustration of RIS-assisted ISAC network is presented in Fig. \ref{fig:RIS-ISAC}.

To the best of our knowledge, the first contribution to secure RIS-assisted ISAC systems is claimed by \textit{Fang et al.}  \cite{9685487}. They optimize SINR in a MIMO radar system with RIS assistance while mitigating eavesdropping threats. The proposed system integrates radar sensing and secure communication, employing joint optimization of transmit beamforming, AN, and RIS phase shifts. The optimization problem, which is non-convex, is solved using a BCD method, a majorization-minimization (MM) algorithm for phase shifts, and first-order Taylor expansion for beamforming. Simulation results demonstrate the RIS's effectiveness in significantly enhancing SINR, with higher performance observed as RIS elements and transmit power increase. The approach ensures secure communication and robust radar detection, validating the potential of RIS in secure DFRC systems.

Following the previous work, in \cite{9747551}, \textit{Mishra et al.} address limitations in previous work which often assume single Eves, neglect indirect paths via RIS, or oversimplify optimization by focusing only on unicast or broadcast scenarios. To overcome these gaps, the paper considers a multicast DFRC system comprising a MIMO radar, multiple legitimate users, and multiple Eves, with both direct and indirect paths through an RIS. The proposed OptM3Sec framework jointly optimizes transmit pre-coding matrices for information and AN along with RIS phase shifts to maximize secrecy rates while ensuring adequate SINR for radar detection and adhering to power constraints. The optimization problem, which is highly non-convex, is solved using BCD and stochastic gradient ascent to handle the complexities of the system model. 

Following \cite{9685487}, \textit{Zhang et al.}  \cite{zhang2023irs} investigate a RIS-assisted DFRC system involving a BS communicating with multiple legitimate users while simultaneously detecting a malicious target, acting as a potential Eve. Three main optimization problems are considered: minimizing Eve's achievable rate, maximizing the minimum SINR for CUs, and maximizing the overall communication sum rate. To address these non-convex optimization problems, the authors use advanced optimization techniques such as FP, Lagrangian dual transformation, and AO.

 Following the previous work, \textit{Liu et al.}  \cite{liu2023drl} address the challenge of maximizing the secrecy rate of legitimate users while mitigating eavesdropping by radar targets, treated as potential Eves. The authors propose a joint optimization strategy for transmit beamforming, AN, and RIS phase-shifts using a deep reinforcement learning (DRL)-based soft actor-critic algorithm. This method tackles the multivariable coupling and non-convex optimization inherent in the problem. The oft actor-critic algorithm incorporates entropy maximization to enhance exploration and avoid local optima. Simulations demonstrate that the proposed approach outperforms benchmarks, including systems without AN or random RIS phase-shifts, in secrecy rate and radar performance.

Next, \textit{Li et al.}  \cite{ li2023irs} address the limitations in previous designs of RIS-assisted secure DFRC systems, which either ignored the complex fourth-order radar SNR or did not effectively convexify it for optimization. The main problem is to maximize the secrecy rate while maintaining radar performance, facing challenges of non-convex objectives and unit-modulus constraints on RIS parameters. The proposed solution employs FP to simplify the secrecy rate into a tractable form and uses minorization techniques to handle the high-order SNR terms. An AO algorithm is developed to jointly design the waveform, AN, and RIS parameters. This approach overcomes earlier limitations by incorporating AN design and addressing the complex interdependencies of the system.

Further, \textit{Salem et al.}  \cite{9973319} propose an \textbf{active RIS}-assisted\footnote{Unlike passive RIS, active RIS can actively boost signal power.} multi-user MISO ISAC system to enhance PLS. The approach involves optimizing transmit beamforming, radar receive beamforming, and the active RIS reflection coefficients to maximize the secrecy rate while meeting radar detection and power budget constraints. FP and MM-based iterative algorithm is developed to solve the non-convex optimization problem.  

Another notable work is done by \textit{Hua et al.} \cite{hua2023secure} where they propose using RIS to enhance the PLS of an ISAC system. In their considered system, the RIS not only assists in downlink communication for multiple users but also creates a virtual line-of-sight link for target sensing.  The study proposes a penalty-based optimization algorithm for scenarios with perfect CSI and a robust algorithm for cases with imperfect CSI and uncertain target location. These methods jointly optimize RIS phase shifts, communication, and radar beamformers to maximize sensing beampattern gain while limiting information leakage. 

 \textit{Wang et al.}  \cite{wang2023star} propose a PLS enhancement for DFRC systems using \textbf{simultaneously transmitting and RIS (STAR-RIS)}\footnote{A special type of RIS that allows both transmission and reflection, offering a 360-degree service coverage.}. The system uses joint waveform design, reflective, and active beamforming to deceive malicious radar targets, preventing potential eavesdropping. The proposed joint optimization approach maximizes radar sensing power while ensuring secure information transmission and quality of service for legitimate users. A novel distance-majorization-based algorithm addresses the nonconvex optimization challenges, providing efficient solutions with reduced computational complexity. 

 In another work \cite{sun2023security} by \textit{Sun et al.} present a security-enhanced ISAC system using phase-coupled \textbf{intelligent omni-surfaces (IOS)}\footnote{A special type of RIS that operates on both sides of RIS  \cite{9491943}. }, which simultaneously supports communication and sensing without extra sensors. The IOS splits space into a communication and sensing part, offering secure communication services and a virtual LOS for target sensing. To prevent information leakage to the sensing target, a joint design of communication and sensing beamformers, alongside IOS phase-shift matrices, is optimized to maximize sensing gain while controlling SINR for secure communication. Two alternative algorithms-independent and coupled-phase models were proposed to optimize the IOS phase shifts effectively.

Next, \textit{Yu et al.}  \cite{yu2024security} investigate enhancing PLS in ISAC systems by utilizing \textit{RIS mounted on an unmanned aerial vehicle (UAV)}. The proposed method introduces AN to disrupt potential eavesdropping from aerial targets while optimizing UAV deployment, beamforming, and phase shifts of the RIS.

\textit{Song et al.}  \cite{song2024secure} propose a secure symbol-level pre-coding design for RIS-aided ISAC systems. The approach involves maximizing the user's SNR while ensuring the CRB for sensing performance, under constraints related to power, constructive interference, and security against Eves. Symbol-level pre-coding is used to enhance communication performance, with RIS enabling additional degrees of freedom by establishing non-line-of-sight links. The non-convex optimization problem is addressed using Taylor expansion and a successive lower bound maximization (SLM) method, leading to improved security and performance compared to systems without RIS. 

Recently, \textit{Jiang et al }  \cite{10561466} consider RIS-NOMA-ISAC systems consists of a BS that serves multiple users while performing sensing tasks, with the RIS introduced to enhance signal quality and mitigate eavesdropping threats. A major challenge arises from NOMA's inter-user interference and vulnerability to eavesdroppers, making secure communication a critical concern. To address this, the authors formulate an optimization problem that aims to maximize the secrecy rate while maintaining the ISAC system's performance requirements. Their proposed solution involves joint active and passive beamforming, where the BS optimizes its transmit beamforming while the RIS adjusts phase shifts to minimize signal leakage to eavesdroppers. Since the problem is highly non-convex, AO techniques and iterative algorithms are employed to achieve a sub-optimal yet efficient solution.

Next, \textit{Yang et al}  \cite{yang2024secure},
 investigate secure transmission in active RIS-assisted ISAC systems, addressing the challenge of eavesdropping on communication signals by malicious entities. Unlike traditional passive RIS systems, active RIS introduces amplifiers to enhance signal manipulation and improve communication reliability and security. The proposed system employs a novel SIC scheme that uses sensing signals as jamming to disrupt Eves while ensuring legitimate communication remains intact. The core problem involves maximizing the secure communication rate by jointly optimizing active RIS beamforming, transmit beamforming for legitimate users and Eves, and radar receive filters under non-convex constraints. The authors propose an iterative optimization algorithm using MM and SDP to decompose the problem into manageable subproblems.

Moreover, in a pre-print \cite{li2024low}, \textit{Li et al.} present a low-complexity design for enhancing PLS in an RIS-assisted DFRC system. The approach integrates RIS to mitigate path loss or blockage, thus increasing the degrees of freedom for optimizing system design. AN is embedded into the radar probing waveform to safeguard communication against eavesdropping targets. The proposed system utilizes AO to design the RIS parameters, radar waveform, and AN, aiming to maximize the secrecy rate while meeting radar SNR constraints. A fractional programming technique and a novel closed-form expression are employed to address the non-convexity of the problem, resulting in a more scalable and computationally efficient solution.  

In another pre-print \cite{liu2024enhancingrobustnesssecurityisac}, \textit{Liu et al.} propose a novel approach to enhance robustness and security in ISAC by leveraging a \textbf{transmissive RIS} (aka STAR-RIS) transceiver and rate-splitting multiple access. The study addresses the challenges of imperfect CSI and models interference to optimize secure beamforming. A BCD-based second-order cone programming (SOCP) algorithm is proposed to solve the nonconvex optimization problem, decoupling the common stream beamforming, private stream beamforming, and time-slot allocation.

\begin{table*}[htb!]
\centering
\caption{Tabular Summary \& Comparison of PLS-based Secure RIS-assisted ISAC Systems}
\begin{tabular}{|p{3cm}|p{4cm}|p{3cm}|p{6cm}|}
\hline
\textbf{References} & \textbf{System Setup} & \textbf{Optimization Techniques} & \textbf{Features} \\ \hline
Fang et al. (2022) \cite{9685487} & MIMO Radar with RIS assistance, Single Target as Eve and a Multi-Antenna CU & BCD, MM, Taylor expansion &  Joint transmit beamforming and RIS phase shift optimization for secrecy rate maximization \\ \hline
Mishra et al. (2022) \cite{9747551} & Multicast DFRC with RIS, Multi-Antenna Multiple CUS, and Mulitple Targets as Eves & BCD, Stochastic gradient ascent &  Robust secure multicast beamforming and RIS design for multiple communication users and eavesdroppers \\ \hline
Zhang et al. (2023) \cite{zhang2023irs} & RIS-assisted DFRC with a Target as Eve, Single Antenna CUs & FP, Lagrangian dual, AO & Secrecy rate maximization through joint beamforming and RIS phase adjustment \\ \hline
Liu et al. (2023)  \cite{liu2023drl}& RIS-aided ISAC with Single Antenna Multiple CUs and a Target as Eve & DRL-based soft actor-critic algorithm & Deep reinforcement learning-based secure beamforming and sensing optimization  \\ \hline
Li et al. (2023) \cite{ li2023irs}& RIS-assisted DFRC with an NLOS Target as Eve, and a Single Antenna Single CU & FP, Minorization techniques, AO & Secure beamforming and RIS design for NLOS target environment using fractional programming\\ \hline
Salem et al. (2023) \cite{9973319} & Active RIS-aided ISAC with Single Antenna Multiple CUs, and a Single Target as Eve & FP, MM-based iterative algorithm &  Active RIS-based joint waveform and phase shift optimization for secure ISAC \\ \hline
Hua et al. (2023) \cite{hua2023secure} & RIS-aided ISAC with NLOS Target as Eve and Single Antenna Multiple CUs & Penalty-based optimization, Robust design for imperfect CSI &  Robust beamforming and RIS phase optimization under imperfect CSI conditions \\ \hline
Wang et al. (2023) \cite{wang2023star} & STAR-RIS-aided DFRC with Single Antenna Multiple CUs and Multiple Targets as Eves  & Distance-majorization-based algorithm &   Joint STAR-RIS configuration and secure beamforming design for ISAC with multiple eavesdroppers \\ \hline
Sun et al. (2023) \cite{sun2023security} & IOS-aided ISAC with Single Antenna Multiple CUs and Targets as Eves & Independent/coupled-phase optimization algorithms &  Secure beamforming and IOS phase design through independent and coupled optimization schemes \\ \hline
Yu et al. (2024)  \cite{yu2024security} & RIS-UAV-aided ISAC with Single Antenna Multiple CUs and a Target as Eve & Joint optimization of UAV deployment, AN, and RIS phase shifts & Joint UAV positioning, artificial noise, and RIS phase optimization for secure ISAC \\ \hline
Song et al. (2024) \cite{song2024secure} & RIS-aided ISAC with Single Antenna Multiple CUs and a Target as Eve & Taylor expansion, SLM method & Low-complexity joint transmit beamforming and RIS optimization for secrecy enhancement  \\ \hline
Jiang et al. (2024) \cite{10561466}  & RIS-NOMA-ISAC with Single Antenna Multiple CUs and a Target as Eve & BCD, AO &  Joint secure beamforming and RIS configuration in RIS-NOMA ISAC systems \\ \hline
Yang et al. (2024) \cite{yang2024secure} & Active RIS-assisted ISAC with Single Antenna CU, Eve and a Target & MM, SDP & Active RIS-assisted secure transmit beamforming and sensing waveform design \\ \hline
Li et al. (2024) \cite{li2024low}  & Low-complexity RIS-aided DFRC with a Single Antenna CU and NLOS Target as Eve and  & FP, AO & Low-complexity secure beamforming and RIS optimization for NLOS-targeted ISAC \\ \hline
Liu et al. (2024) \cite{liu2024enhancingrobustnesssecurityisac} & TRIS-aided ISAC with Single Antenna Multiple CUs and Targets as Eves & BCD, SOCP & Robust transmit beamforming and TRIS phase optimization for secure ISAC with multiple eavesdroppers \\ \hline
\end{tabular}
\label{tab:ris_isac_comparison}
\end{table*}

\subsubsection{NTC-assisted ISAC}
\begin{figure*}
    \centering
    \includegraphics[width=1\linewidth]{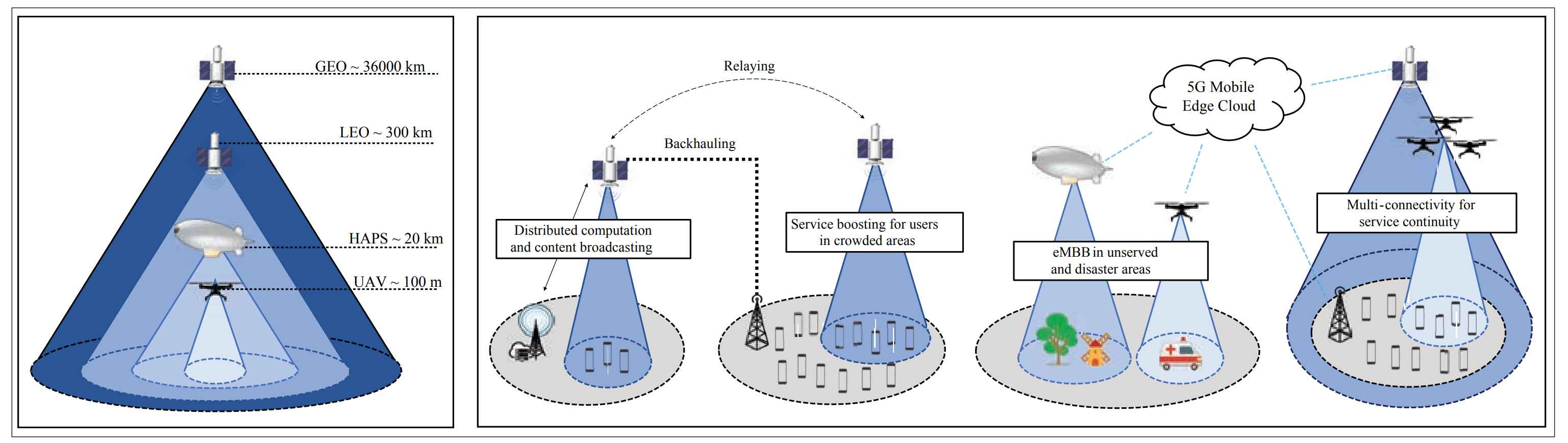}
    \caption{Overview of NTNs showing hierarchical communication from geostationary earth orbit (GEO), LEO, HAPS, and UAVs, supporting applications like distributed computation, service boosting in crowded areas, enhanced mobile broadband (eMBB) in disaster zones, and multi-connectivity for service continuity, integrated with a 5G mobile edge cloud  \cite{9275613}.}
    \label{fig:NTN}
\end{figure*}

NTC refers to communication systems that operate through platforms other than traditional terrestrial networks, such as satellite, aerial, or space-based systems. These platforms provide communication services beyond the limitations of ground-based infrastructure, offering broad coverage, connectivity in remote areas, and resilience against natural disasters or man-made disruptions \cite{azari2022evolution}. The key systems of NTC are satellite communication (Satcom), high-altitude platform systems (HAPS), UAVs, and space-based communication \cite{9193893}. An illustration of NTN is presented in Fig. \ref{fig:NTN}.
Low-earth orbit (LEO) satellites play a significant role in enhancing availability in unserved and under-served areas such as deserts, oceans, and rural areas to achieve the goal of global ubiquitous and high-capacity connectivity. Due to the broadcast
nature and extensive satellite communications coverage, satellite communication information is susceptible to eavesdropping
threats, making security a crucial concern.

\begin{figure}
    \centering
    \includegraphics[width=\linewidth]{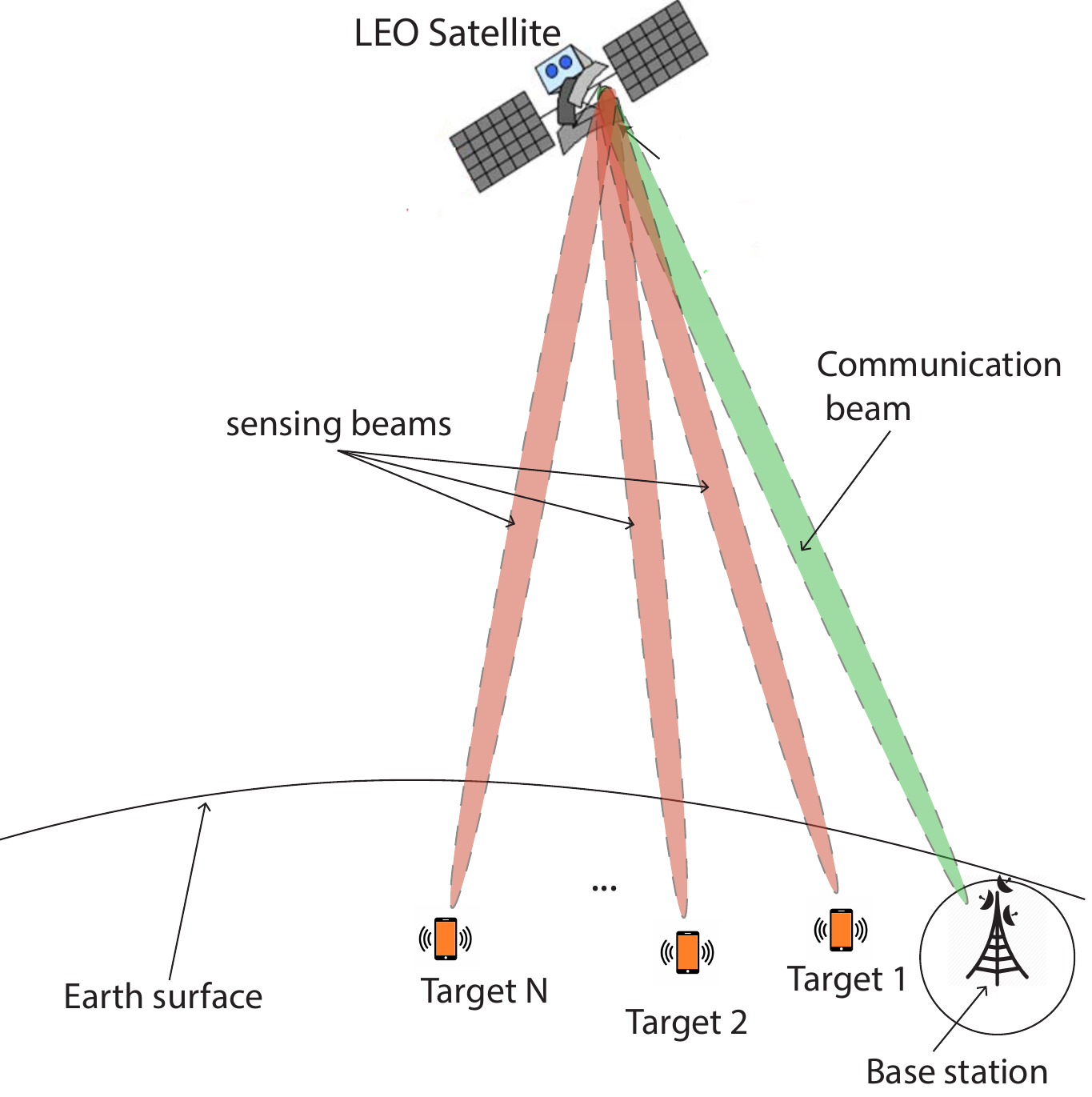}
    \caption{{An illustration of LEO satellite assisted ISAC where satellite uses communication and sensing beams to interact with a BS and detect multiple targets on the Earth's surface.}}
    \label{fig:enter-label}
\end{figure}

\begin{figure}
    \centering
    \includegraphics[width=1\linewidth]{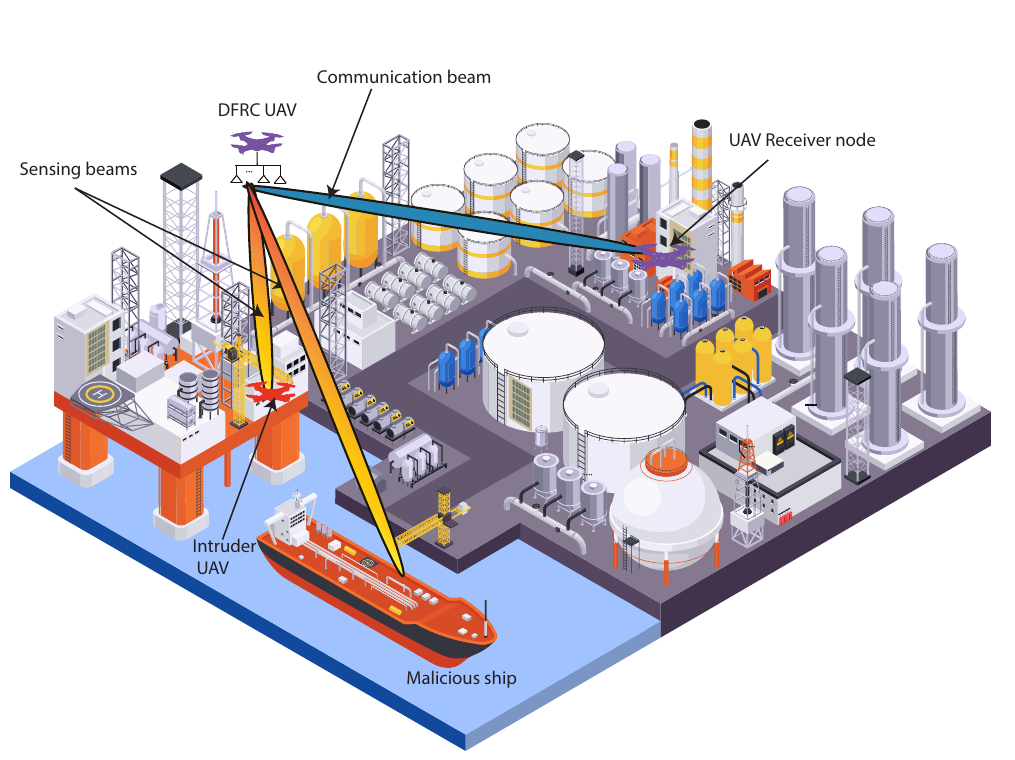}
    \caption{{An illustration of a DFRC UAV employing communication and sensing beams to monitor and communicate with a receiver node while identifying an intruder UAV and a malicious ship.}}
    \label{fig:enter-label}
\end{figure}

To the best of our knowledge, the first attempt on securing ISAC-enabled LEO satellite systems is done by \textit{Zhang et al. }  \cite{zhang2024joint}. By leveraging the satellite's sensing capability, the system can estimate the angle of potential Eves to enhance security. The main objective is to maximize the worst-case sum secrecy rate through joint optimization of secure transmit beamforming and radar receive filters under sensing and power constraints. The authors use optimization techniques like the concave-convex procedure and Taylor series expansion to solve the resulting non-convex problem iteratively. 

On the other hand, considering secure UAV-enabled ISAC, we found the first notable work \cite{wu2023uavs} by \textit{Wu et al.}. They focus on the real-time trajectory design for UAV-aided ISAC systems, addressing challenges in secure communication. The authors aim to maximize the real-time secrecy rate while tracking a legitimate user with unknown movements and mitigating data leakage to an Eve on a known trajectory. To solve this, the authors propose an extended Kalman filter (EKF)-based method to track and predict legitimate user location using sensing echoes. The trajectory optimization problem, which is non-convex due to coupling between variables, is tackled with an iterative algorithm leveraging SCA. Simulation results demonstrate that the proposed algorithm efficiently tracks legitimate user movements and achieves a balance between legitimate communication and data leakage reduction. The study highlights UAV-ISAC systems' potential for secure communication in dynamic environments, with flexible trajectory designs adapting to diverse application. 

In the same year, another notable work \cite{10345500} by \textit{Wei et al.},  proposes an integrated sensing, navigation, and communication framework to secure UAV-enabled wireless networks against mobile Eves. The proposed framework addresses the mobility challenge of Eve by combining AN for jamming and sensing, enabling the information UAV to estimate the Eve UAV's state using an EKF. This information guides real-time navigation, wiretap channel prediction, and resource allocation to secure communication and optimize tracking performance. The framework employs a divide-and-conquer approach for navigation and a neural network-based method to model wiretap channel prediction errors.

Following the previous work \cite{wu2023uavs},  \textit{Xiu et al.}  \cite{Xiu:TVT:2024}  focus on enhancing security for a multiple-antenna UAV that transmits ISAC waveforms to communicate with several ground IoT devices while also detecting its surroundings. They maximize the average communication secrecy rate by optimizing both the UAV's trajectory and beamforming vectors. Since the UAV trajectory optimization problem is non-convex due to the coupling of multiple variables, the authors develop an efficient algorithm based on SCA.

Following the previous work \cite{wu2023uavs}, \textit{Liu et al.}  \cite{Liu:TVT:2024} consider multiple Eves. They introduce a dual-UAV system, where the source UAV performs sensing and communication while a jammer UAV emits jamming signals to disrupt Eves. The key challenge lies in jointly optimizing user scheduling, transmit power, and the source UAV's trajectory to maximize the secure transmission rate under non-convex constraints. The authors decompose the optimization task into three subproblems, iteratively solving them using SCA and relaxation techniques. The proposed solution ensures that legitimate users receive optimal service while Eves are effectively disrupted.

In the same year, \textit{Li et al.}  \cite{li2024uav} propose a PLS strategy for a multi-UAV-assisted ISAC  network using a Stackelberg game model. The study considers a dual-function BS communicating with ground users while actively countering an Eve. The Stackelberg game optimizes the transmit power strategies between UAVs and Eve, considering both perfect and imperfect CSIs. The CRB is used to model the sensing accuracy of Eve's location, which impacts the game dynamics. The proposed iterative algorithm seeks a suboptimal solution for both communication and sensing operations, ensuring security and efficient sensing. 


Recently, in a pre-print, \textit{Son et al.} \cite{son2024secrecy} introduce a UAV-enabled ISAC  system, where a \textit{FD-UAV} equipped with a uniform planar array functions as a BS for multiuser downlink communications while also sensing and jamming a passive ground Eve. They maximize the sum secrecy rate of ground users, considering constraints related to sensing accuracy and the UAV's operational capabilities. To achieve this, the authors propose a joint optimization approach for transceiver beamforming and the UAV's trajectory. They develop an algorithmic solution utilizing BCD and SDR techniques.  

In another pre-print \cite{xiu2024secure}, \textit{Xiu et al.} analyze the security performance of a \textit{RIS-aided UAV} communication system that integrates sensing and communications. They focus on a multiple-antenna UAV that transmits ISAC waveforms to simultaneously detect an untrusted target and communicate with a ground  IoT device, all while accounting for the presence of an Eve. Recognizing that Eve may conceal their CSI in real-world scenarios, the study assumes imperfect CSI for the Eve channel. The objective is to maximize the average communication secrecy rate by jointly optimizing the UAV's trajectory, RIS passive beamforming, transmit beamforming, and receive beamforming. Due to the non-convex nature of this joint optimization problem caused by multi-variable coupling, the authors propose an efficient algorithm using BCD methods. They develop a SCA algorithm based on SDR to tackle the problem by breaking it into four separate non-convex subproblems.

\begin{table*}[htb!]
\centering
\caption{Tabular Summary \& Comparison of PLS-based Secure Non-terrestrial ISAC Systems}
\begin{tabular}{|p{3cm}|p{4cm}|p{3cm}|p{6cm}|}
\hline
\textbf{References} & \textbf{System Setup} & \textbf{Optimization Techniques} & \textbf{Features} \\ \hline
Zhang et al. (2024) \cite{zhang2024joint} & LEO Satellite ISAC with Single Antenna Multiple CUs and a Target as Eve & Concave-convex procedure, Taylor expansion & Joint secure beamforming and sensing waveform design for LEO satellite ISAC\\ \hline
Wu et al. (2023) \cite{wu2023uavs} & UAV-aided ISAC with Single Antenna CU as Target and an Eve & EKF, SCA, Iterative algorithm & Secure trajectory and beamforming optimization for UAV-aided ISAC systems\\ \hline
Wei et al. (2023) \cite{10345500}  & UAV-enabled ISAC with Single Antenna CUs and a UAV Target as Eve & Divide-and-conquer, Neural network-based method & UAV trajectory and secure beamforming design using neural network-based optimization \\ \hline
Xiu et al. (2024) \cite{Xiu:TVT:2024} &UAV-enabled ISAC with a Single Antenna CU and a Target as Eve & SCA &Secure communication and sensing beamforming design through successive convex approximation\\ \hline
Liu et al. (2024) \cite{Liu:TVT:2024} & UAV-enabled ISAC with a Jammer UAV, and Singel Antenna CUs and Targets as Eves & SCA, Relaxation techniques &Joint UAV trajectory optimization and secure communication-sensing beamforming with jammer presence  \\ \hline
Li et al. (2024) \cite{li2024uav} & Multi-UAV-assisted ISAC with Single Antenna CUs and a Target as Eve & Stackelberg game, Iterative algorithm & Stackelberg game-theoretic secure resource allocation for multi-UAV-assisted ISAC\\ \hline
Son et al. (2024) \cite{son2024secrecy} & FD-UAV ISAC with Single Antenna CUs and a Target as Eve & BCD, SDR & Full-duplex UAV secure beamforming and sensing waveform design for secrecy maximization\\ \hline
    Xiu et al. (2024) \cite{xiu2024secure} & RIS-aided UAV ISAC with Single Antenna CU and Eve, and Sensing Target as Eve  & BCD, SCA, SDR & Joint RIS configuration and secure beamforming design  \\ \hline
\end{tabular}
\label{tab:uav_leo_isac_comparison}
\end{table*}

\subsection{{PLS for Covert Communication}}
{
So far, we have explored ISAC's potential in enabling secure transmissions and the existing secrecy-sensing trade-off in ISAC-enabled wireless networks from a communication confidentiality point of view, generally quantified in terms of the secrecy capacity. However, communication covertness is a another wireless communication security that recently attracted a significant amount of interest. Herein, the objective is to obfuscate the communication from the Eves' (also known as wardens) reach and confuse their observations. In other words, the core aim is to confuse the wardens on whether a transmission is taking place or not. Hence, such an aim is achieved by the wise optimization on the system's resources, such as the transmit power, beamforming pattern, RIS reflection pattern, and legitimate transceivers' locations. However, despite the aforementioned enabling means for reaching covert communications, several challenges impede it, mainly the absence of information
In \cite{covisac1}, the authors proposed a radar sensing-based scheme, where a separate radar sensor, coexisting and cooperating with the transmitting BS to locate an aerial warden Willie $(W)$ and estimating its corresponding CSI with respect to the transmitting BS. The estimated which will be used for maximizing the covert communication rate. The estimated CSI of $W$ (sensed target) from the sensing parameters (i.e., range and azimuth/elevation angles) is used for optimizing the BS beamformer. Furthermore, an EKF is used for tracking the moving $W$ in real time and refining the beamforming design. In \cite{covisac2}, the authors extended the covert communication analysis in ISAC networks by considering several suspicious wardens (targets). Under the consideration of two different types of CSI imperfection for the transmitter-wardens' channels, an optimal beamforming and AN injection design is proposed to maximize the sensing performance subject to covertness and reliability constraints.
The authors of \cite{covisac3} dealt with the performance optimization of an ISAC-enabled wireless covert communication network. In particular, the target detection mutual information metric is maximized subject to covert rate and power constraints. It is worth mentioning that the considered setup consisted of separate Willie and target nodes, whereby authors inspected as well the performance limits of the proposed scheme in the presence of imperfect CSI of the channel with respect to Willie. In \cite{covisac4}, a similar setup to the previous work was analyzed by optimizing the transmit communication and sensing beamforming weights minimizing the sensing CRB, subject to a minimal legitimate SINR and a maximal covertness entropy. The work of Hu \textit{et al.} in \cite{covisac5} incorporated a RIS to enhance communication covertness in an ISAC-enabled covert communication network. The considered scheme aimed at optimizing the RIS phase shifts to increase the legitimate user's communication rate, subject to a certain covertness level with respect to Willie. It is worth mentioning that such a scheme considered the RIS to not affect the sensing link with respect to the sensed flying target.}

\subsection{PLS for Authentication}
\begin{figure*}
         \centering
         \includegraphics[scale=.05]{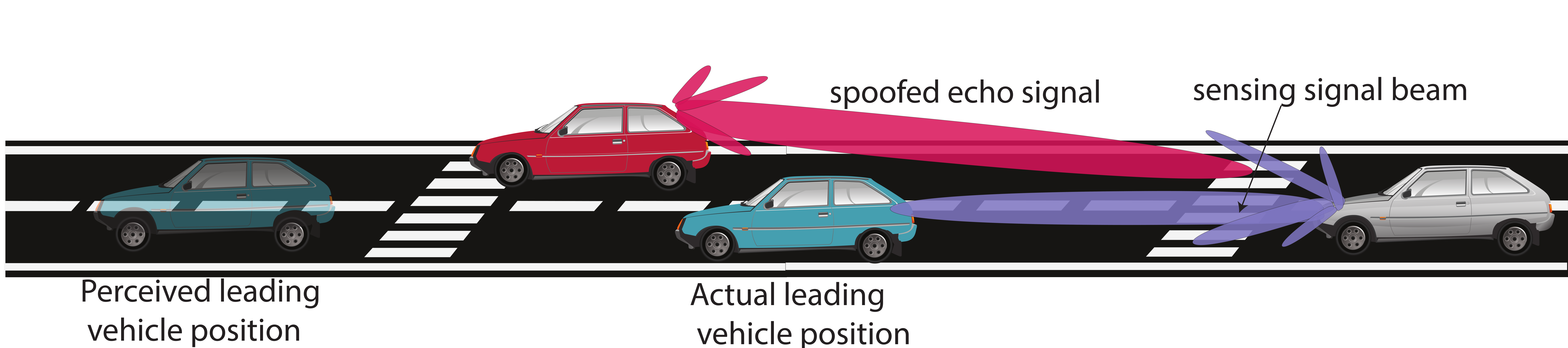}
         \caption{ISAC-enabled vehicle-to-vehicle communication case under sensing spoofing attacks.}
         \label{plafig1}
     \end{figure*}

\begin{figure*}
         \centering
         \includegraphics[scale=.45]{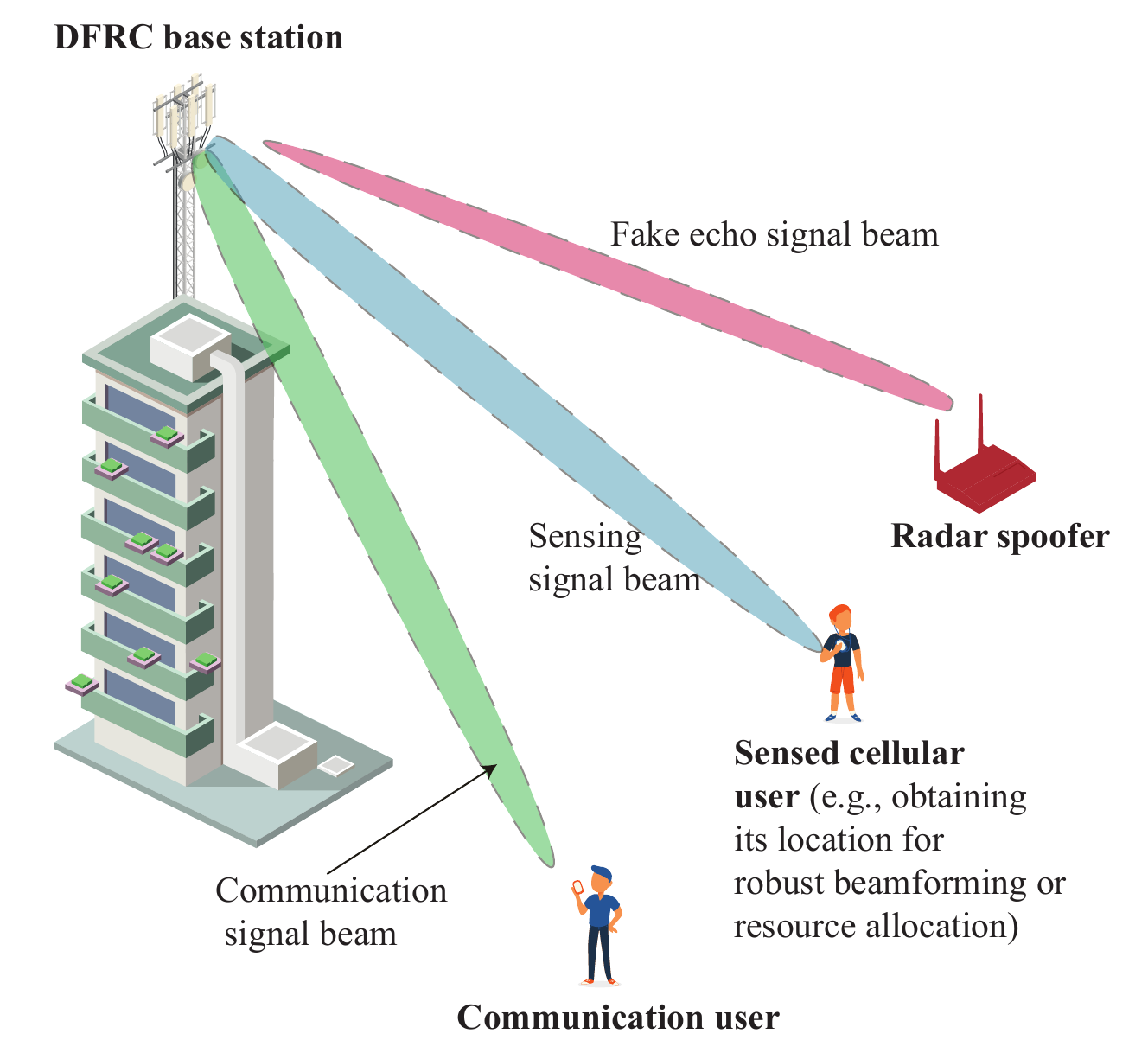}
         \caption{ISAC-enabled cellular network case under sensing spoofing attacks.}
         \label{plafig2}
     \end{figure*}

In wireless networks' security, the authentication process is tasked with verifying the identity of the sender to prevent spoofing, impersonation, or data manipulation (e.g., man-in-the-middle) attacks. In the context of ISAC networks, authentication gets an additional degree of cruciality due to the availability of another attack surface: Sensing information. In this aspect, attackers might not only target impersonating a node's identity or altering a data packet but also aim at spoofing/faking a sensing signal in order, for instance, to provoke an incorrect environment mapping (e.g., an intruder in a natural resource facility trying to hide itself from being sensed by altering the radar echo signals) or alter the measured physical characteristics of a sensed object from the received echo signal (e.g., speed, direction). Figs.\ref{plafig1} and \ref{plafig2}  showcase potential applicability scenarios for sensing spoofing.

From a PLS point of view, PLA, which is a subcategory of PLS, aims at leveraging PHY features to enable node or message authentication. However, it should be emphasized that a scarce amount of work tackled sensing spoofing attacks in and/or proposed respective PHY-assisted schemes for attack detection. Specifically, the surveyed techniques, detailed in the sequel, were proposed in the context of standalone sensing networks (e.g., radar sensing, wireless sensors network), while, to the best of the authors' knowledge, the coverage of such a security pillar in ISAC has not yet been considered. Nonetheless, the following review of PLA papers in the context of standalone sensing-based communication systems can stimulate insightful ideas in their involvement in the context of ISAC and exploring spoofing/manipulation attacks on the node identity, message, or sensing signal.

In \cite{auth1}, the authors analyzed the impact of spoofing attacks in a mmWave-based radar sensing network, where the malicious nodes aim at faking and/or replaying the genuine echo signal by altering the physical properties of the sensed object/targets (e.g., velocity, distance) embedded in the signal. By means of an experimental vehicular network testbed, the vulnerability level of radar spoofing over various cases is assessed. Furthermore, as a solution to this type of attack, a (i) challenge-response technique, based on embedding randomly chosen parameters on the transmit pulses (e.g., initial phase, frequency) and (ii) a signal statistics-based radio frequency fingerprinting (RFF) were proposed and evaluated. Likewise, the authors in \cite{auth2} proposed a PHY challenge-response scheme to detect sensing spoofing in wireless sensor networks. The proposed scheme is based on embedding a challenge, represented by a null-amplitude interval, on the transmitted sensing. Such an approach can help in detecting hostile spoofing transmissions by malicious entities aiming to harm the environment's sensing performance by the sensor. Also, Kapoor \textit{et al.} in \cite{auth3} proposed a novel spoofing attack detection scheme in MIMO radar systems applied in a vehicular network, by exploiting the angular beam directivity and the radar transmit and receive MIMO arrays. Spoofing attacks are efficiently spotted by randomizing the narrow beam selection as well as illuminating the area ahead of a sensing vehicle with multiple beams. By inspecting the received echo signal's properties, malicious fake echo signals can be detected. In \cite{auth4}, the authors proposed a frequency hopping-based scheme to avoid radar spoofing in a frequency-modulated continuous-wave radar scheme employing chirp-spread spectrum signaling. From another front, some works aimed at detecting radar jamming attacks, whereby some of the radar spoofing detection mechanisms are also applicable for jamming detection. For instance, in \cite{auth5}, the authors proposed a spectral analysis-based jamming detection scheme by analyzing the measured Fourier spectrum of the received echo signal over several spectral windows and comparing the number of evaluated median values to a fixed threshold.

\section{{Current Challenges and} Future Research Directions}
{Despite the achieved research status in the development of ISAC-enabled networks, particularly in improving the secrecy of such networks, several challenges are still impeding the integration of various enabling techniques in securing ISAC networks. We should highlight that while some of these challenges are purely related to the nature of ISAC systems, others are inherited from the PLS paradigm irrespective of the considered framework.}

\subsection{{Finite Blocklength Regime: A Revisited Information-Theoretic Security Limitation}}

{Sensing is expected to form a major pillar in 6G-driven use-cases and applications, where a centimeter-level sensing is targeted. Furthermore, similarly to the current 5G networks,  several 6G applications are low-latency-driven by transmitting short data packets. One of the critical theoretical assumptions in secrecy-capacity-based PLS is the infinite blocklength consideration, where the codeword is assumed to be infinitely long, yielding a decoding error probability tightly approaching zero and the communication rate achieving the Shannon capacity. Consequently, such an infinite blocklength consideration cannot be applied to the design and evaluation of secure ISAC networks, due to the actual finite blocklength (i.e., packet/codeword length) in several 5G/6G classes of service. As far as the proposed secure ISAC designs are concerned, the vast majority of them consider an incompatible infinite blocklength assumption with short-packet low-latency transmissions. Therefore, it is of paramount importance to quantify the theoretical secrecy limits of ISAC network considering a finite blocklength regime, in which the code rate and redundancy rate are optimized to balance the secrecy-latency trade-off. In such an instance, sensing constraints should also be considered in the system design problem.}

\subsection{{Manipulative Attackers: Additional Security Threat Degrees}}

{One of the main limitations in PLS is the absence of information about the eavesdropping channels to the passive Eve's nature. In this context, sensing can play a fundamental role in localizing the Eve and constructing an estimate for the wiretap channel response at the BS, which can help in evaluating and optimizing the considered network's secrecy. Nonetheless, a smart malicious entity can aim for node damage by disrupting the sensing process, e.g., altering/modifying the echo signal power, injecting a harmful interfering jamming with the radar echo. Such active attacking procedures will target provoking an incorrect estimate of the Eve's location, and consequently its channel, which can subsequently favoritize legitimate signal's beamforming by the transmitter towards the Eve(s). One potential way to address such an issue is the use of several multistatic radar receivers to improve the accuracy of localization and evade malicious radar jamming. Therefore, the development of robust and flexible sensing mechanisms that can combat such active attacks on sensing stands out as a crucial open research problem in secure ISAC networks.}

\subsection{{Sensing Eavesdropping}}

{In wireless security, the \textit{de facto} objective is preserving the legitimate information out of the reach of the attacker nodes or elaborate solid shielding mechanisms against impersonation and spoofing attacks. However, one can project several of these wireless security menaces in the context of sensing, where the attacker may be interested in haunting information regarding the sensed environment/targets. In such an instance, malicious nodes can try exploiting the omnidirectional nature of reflected echo signals from the sensed targets in the environment and the partial information about the radar transceivers to estimate sensitive localization/range parameters, i.e., distance to the target, angle of arrival, speed, etc. Such a type of attacks is currently overlooked in the realm of secure ISAC networks design, where, by analogy to the information-theoretical limits achieved/evaluated with respect to the legitimate data information (i.e., Wyner's secrecy capacity), an equivalent measure should be taken with respect to the achievable sensing mutual information at the evesdropper. Therein, several methods need to be applied, such as signal obfuscation methods.}

\subsection{PLS in Emerging Technologies in ISAC}
  As wireless communication systems evolve towards 6G and beyond, the push for higher data rates, ultra-low latency, and seamless connectivity brings the THz spectrum (0.1-10 THz) to the forefront due to its promise of supporting ultra-high bandwidth. Concurrently, integrated sensing functions add significant value by enhancing spatial awareness, which is crucial for applications such as autonomous driving, augmented reality, and smart environments \cite{han2024thz}. One of the primary challenges of using THz frequencies is the severe propagation loss due to high free-space path loss and atmospheric absorption. This limits the coverage of THz communication systems, particularly in outdoor environments where humidity can have a significant impact \cite{elbir2024terahertz}. To overcome this, line-of-sight (LoS) transmission and RIS are often used, but these add system complexity.
  These channels are susceptible to eavesdropping due to the narrow beams, which could be intercepted if alignment is not precise. PLS techniques such as secure beamforming, jamming, and AN injection need to be further explored in the THz band to ensure the security and privacy of ISAC operations.

On the other hand, orthogonal time frequency space (OTFS), proposed by Monk et al., is a novel modulation scheme that is particularly well-suited for high-mobility communication scenarios, such as vehicular networks and high-speed trains~\cite{monk2016otfsorthogonaltime}, as well as for environments with significant multipath fading. Unlike traditional modulation schemes like orthogonal frequency division multiplexing (OFDM), OTFS aims to achieve robustness against time-frequency variations in wireless channels, making it a promising candidate for next-generation wireless communication, including 5G and beyond. Recently, an OTFS-ISAC system was investigated in~\cite{10445333}, where the enhanced impact of OTFS was demonstrated through results. Unlike traditional time-frequency-based security methods, OTFS leverages the delay-Doppler domain, naturally enhancing security by spreading symbols across multiple paths, making it difficult for adversaries to intercept or manipulate transmissions. 
However, implementing PLS in OTFS-ISAC faces challenges in delay-Doppler channel estimation, where securing dynamic wireless environments is complex. Besides, there is a trade-off between communication and sensing security, as methods like AN can protect data but degrade sensing accuracy. We believe that a thorough investigation is required in this direction. 

\subsection{{AI-Enabled ISAC: A Friendly Approach?}}
Machine learning (ML) models can be used to adaptively learn the characteristics of the wireless channel in ISAC systems. This helps to improve channel estimation accuracy, allowing ISAC systems to better distinguish between legitimate and malicious signals, which enhances secrecy. ML-based anomaly detection techniques, particularly unsupervised learning algorithms like Autoencoders or clustering, can identify anomalies in communication or sensing data. Such detection helps in identifying malicious activities early and applying necessary countermeasures. ML-based feature extraction techniques can be used to intelligently process the received signals to distinguish between legitimate and malicious activities. For example, convolutional neural networks (CNNs) can be used to extract meaningful features that help in differentiating between authorized and unauthorized users. Deep learning models can be used to determine the best reflection coefficients of the RIS, thereby ensuring that the reflected signals are secure from unintended receivers while maximizing the quality for legitimate users deep reinforcement learning (DRL) can be employed to develop adaptive security strategies. For instance, selecting transmission power levels and beamforming patterns or introducing AN can be optimized using DRL to maximize secrecy capacity under changing channel conditions.
To the best of our understanding, we also highlight some of the challenges in this research direction such as
ML models trained on specific ISAC configurations may struggle to generalize across different system architectures, hardware platforms, or frequency bands. Many ISAC devices, such as IoT sensors, operate with limited computational resources, making it challenging to deploy complex ML models.
\section{Conclusions}
This paper provided a comprehensive survey of physical layer security (PLS) in integrated sensing and communication (ISAC) systems, addressing their dual functionality as a transformative paradigm in future wireless networks, particularly in the 6G era. ISAC systems leverage shared spectral and hardware resources to optimize communication and sensing capabilities, but this integration introduces unique security challenges. To address these challenges, PLS has emerged as a promising approach to enhance data confidentiality, covertness, and resilience against adversarial threats at the PHY.

This survey addressed a critical gap in the existing literature by systematically reviewing state-of-the-art PLS techniques specifically developed for ISAC systems. Unlike prior work that primarily identified security threats or proposed high-level research directions, we provided a comprehensive analysis of recent advancements in the domain of PLS for ISAC. First, we established a foundational understanding of PLS by discussing its core components-confidentiality, covertness, and authentication-along with their mathematical background. We then presented a detailed and system-oriented review of PLS techniques employed in ISAC systems, systematically categorizing key contributions and highlighting their distinctive features through concise tabular comparisons. Finally, we explored emerging and innovative directions for future research, aiming to inspire the development of novel PLS strategies to address the unique challenges posed by ISAC systems.  We identified several open challenges, including finite blocklength security, sensing eavesdropping, and achieving a balance between security and system efficiency. Additionally, we discussed the potential of emerging tools, such as machine learning (ML) and deep reinforcement learning (DRL), to enhance PLS adaptability and robustness in dynamic ISAC environments.

This work aims to provide researchers with a consolidated resource for understanding the current landscape of PLS in ISAC systems, while also inspiring future research to address the identified challenges. As ISAC systems continue to evolve, advancing PLS strategies will be critical to achieving secure, efficient, and resilient communication and sensing capabilities in 6G networks and beyond.

\appendices

\footnotesize{
\bibliographystyle{IEEEtran}
\bibliography{references}
}

\vfill\break

\end{document}